\documentclass[aps,prl,twocolumn,groupedaddress,showpacs]{revtex4}

\usepackage[bookmarks]{hyperref}
\usepackage[dvips]{graphicx}
\usepackage{amsmath}
\usepackage{amstext}
\usepackage{graphics}
\usepackage{exscale}
\usepackage{epsfig}
\usepackage[T1]{fontenc}
\usepackage{bm}
\usepackage{bbm}
\usepackage{color}

\usepackage{version}

\usepackage{latexsym}

\usepackage[tight]{subfigure}

\newcommand{\colortxt}[1]{{#1}}

\renewcommand{\epsilon}{\varepsilon}

\newcommand{\ket}[1]{\! | #1\ \!\! \rangle}

\newcommand{\integral}[3]{\!\int\limits_{#2}^{#3}\!\!{\rm d}#1\;}

\newcommand{\expval}[2]{ \langle  #1 #2\ \!\! \rangle}

\newcommand{\elcre}[2]{ c^{\dagger}_{#1,#2}}
\newcommand{\elann}[2]{ c_{#1,#2}}

\newcommand{\e}{\mathrm e}

\newcommand{\vct}[1]{\bm #1}
\newcommand{\vk}{{\bm k}}

\newcommand{\kF}{k_{{\scriptscriptstyle \mathrm{F}}}}

\newcommand{\epsF}{\epsilon_{{\scriptscriptstyle \mathrm{F}}}}

\newcommand{\HI}{H_{\mathrm{int}}}

\newcommand{\Imag}{\mathrm{Im}}
\newcommand{\Real}{\mathrm{Re}}

\newcommand{\comm}[2]{\left[ #1, #2 \right]}

\newcommand{\hc}{\mathrm{h.c.}}

\includeversion{commentst1} 

\begin{document}

\title{Realizing a Kondo-correlated state with ultracold atoms}  
\author{Johannes Bauer,${}^1$ Christophe Salomon,${}^2$ and Eugene Demler${}^1$}
\affiliation{${}^1$ Department of Physics, Harvard University, Cambridge,
  Massachusetts 02138, USA}
\affiliation{${}^2$ Laboratoire Kastler Brossel, CNRS, UPMC, Ecole Normale
  Superieure, 24 rue Lhomond, 75231 Paris, France} 
\date{\today} 

\begin{abstract}
We propose a novel realization of Kondo physics with ultracold atomic gases. It is
based on a Fermi sea of two different hyperfine states of one atom species
forming bound states with a different species, which is spatially confined
in a trapping potential. We show that different situations
displaying Kondo physics can be realized when 
Feshbach resonances between the species are tuned by a magnetic field and the
trapping frequency is varied. We illustrate that a mixture of
${}^{40}$K and ${}^{23}$Na atoms can be used to generate a Kondo correlated
state and that momentum resolved radio frequency spectroscopy can provide unambiguous
signatures of the formation of Kondo resonances at the Fermi energy. We discuss how
tools of atomic physics can be used to investigate open questions for Kondo
physics, such as the extension of the Kondo screening cloud.  
\end{abstract}

\maketitle

\paragraph{Introduction -} 
Significant advances in quantum optics, such as in cooling, trapping and manipulating
ultracold atoms, have lead to the realization of a plethora
of exciting many-body phases in a controlled manner \cite{BDZ08,HCZDL02}. 
An important drive for this field is that many of these phases and their
description in terms of model Hamiltonians are of great interest in condensed
matter physics, such that a fruitful interplay of these fields has developed. 
It has, for instance,  been possible to realize superfluid phases both in
fermionic and bosonic systems and the transition to a Mott insulating regime
\cite{GMEHB02,GRJ03,ZSSSK05,CMLSSSXK06,Sea08,Jea08}.

An intriguing many-body effect in condensed matter physics is the Kondo
effect. It occurs when itinerant fermions interact with magnetic impurities,
such as, for instance, a small concentration of Fe in Au. The orbital
occupation of the impurity must be such that there is an
unscreened spin present, i.e., in the simplest case a localized state occupied
by a single electron. The essence of the Kondo effect is then that at low
temperature this electron spin forms a many-body bound state with the
itinerant electrons and becomes 
magnetically screened. Crucial for this magnetic screening are second order
processes which lead to frequent spin flips. This Kondo-correlated state
leads to a distinctive 
feature in the resistivity (Kondo minimum) and was also observed as enhanced
transport in quantum dots \cite{GSMAMK98,COK98}. 

In spite of decades of intense research \cite{hewson}, there remain unresolved
questions. For instance, a Kondo screening cloud with a
certain spatial extent and characteristic oscillations was predicted \cite{hewson,AS01,Bor07,ABS08}, 
however, its experimental observation has remained elusive. 
On increasing the impurity concentration from very few to a full lattice, the
Kondo clouds overlap and the localized spins interact
with each other via the so-called Ruderman-Kittel-Kasuya-Yoshida (RKKY)
coupling, mediated by the itinerant 
fermions. This generates a competing effect to the Kondo screening and leads
to a transition to a magnetically ordered state of the
spins. The Kondo lattice problem is of paramount 
importance for the understanding of heavy fermion systems and quantum
criticality \cite{SS10,LRVW07},
however, it is very hard to analyze it theoretically beyond the mean field
level. Here, we propose an experimental setup based on ultracold atoms to
realize single impurity and lattice Kondo situations. The Kondo scale
is shown to be accessible by current experimental techniques. 

\begin{figure}[!t]
\centering
\subfigure[]{\includegraphics[width=0.44\columnwidth]{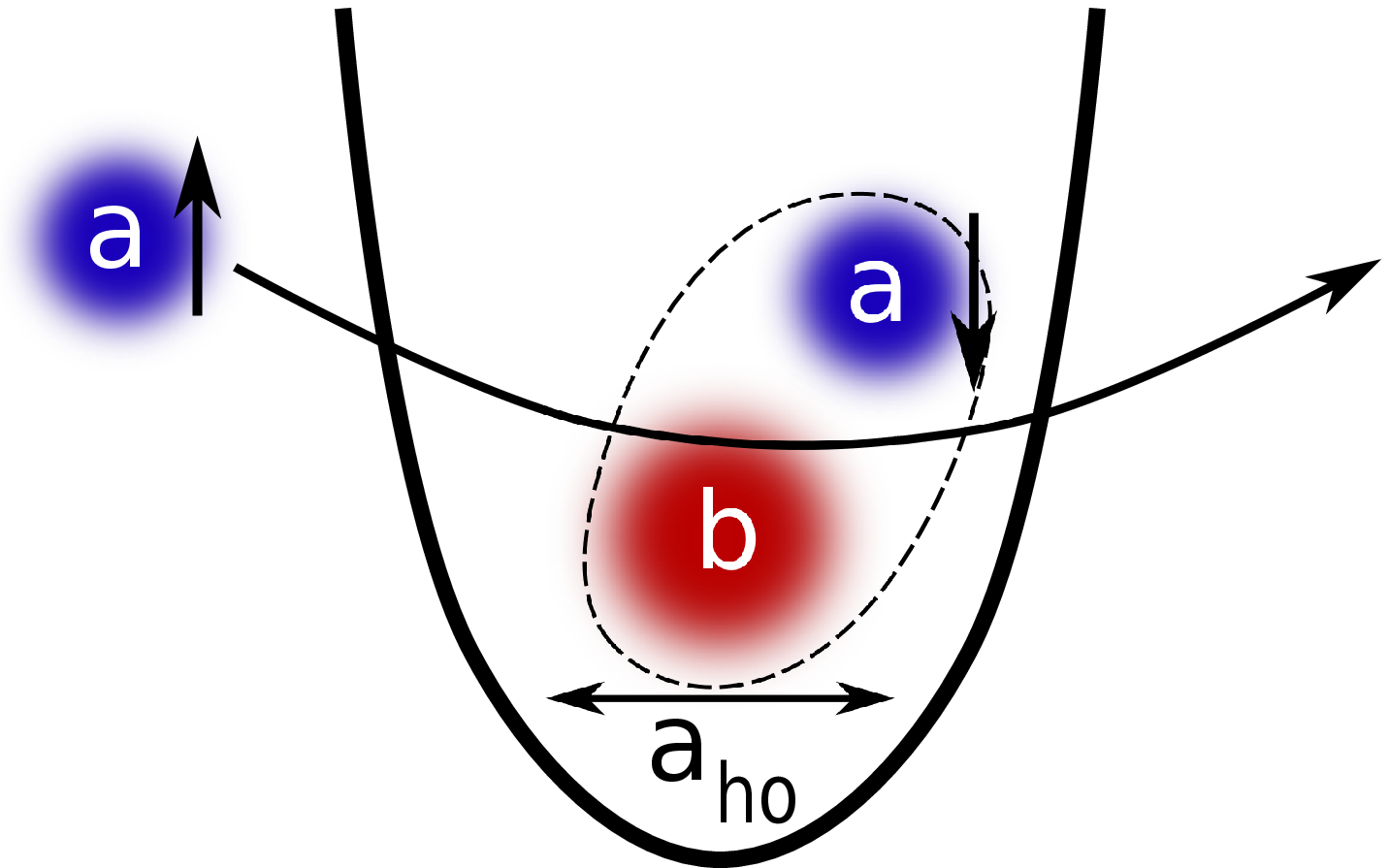}}
\hspace*{0.1cm}
\subfigure[]{\includegraphics[width=0.52\columnwidth]{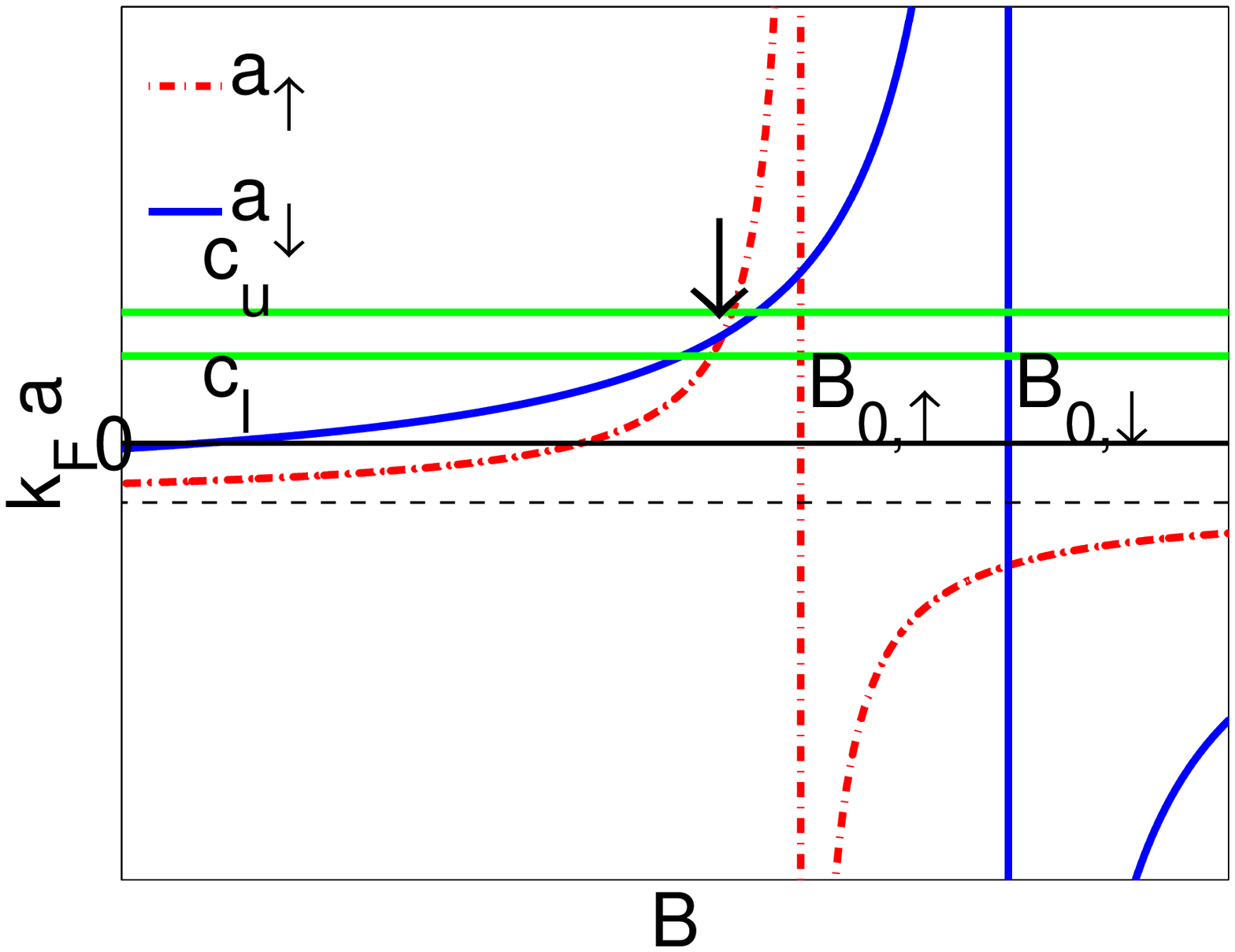}}
\vspace*{-0.3cm}
\caption{(Color online) (a) Schematic picture of an atomic bound state of an $a$
  species atom with a $b$ species atom in a harmonic trap with oscillator
  length $a_{\rm ho}$. (b) Schematic plot  of the effective scattering length
  $a_{\sigma}$ close to a Feshbach resonance for a case where the
  $a_{\uparrow}=a_{\downarrow}$ is satisfied within a suitable regime of
  parameters for Kondo physics given by the boundaries $(c_l,c_u)$ as 
  explained in the text.} 
\label{fig:1}
\vspace*{-0.5cm}
\end{figure}
In the field of ultracold atoms, a lot of progress has recently been made
for manipulating mixtures of different species that offer the unique
possibility to selectively trap one species and to handle heteronuclear
Feshbach resonances producing molecular bound states
\cite{KGJ06,CGJT10,Pea12,WPAWZ12,Tea13,RPUHKWT13}. These 
developments are important ingredients for our proposal. 
The main idea is to allow atoms of a Fermi sea to form bound
states with a different, spatially confined atom, the impurity [see
Fig.~\ref{fig:1}~(a)]. We consider two  
species of ultracold atoms with mass $m_a,m_b$. Species $a$ is   
fermionic and is prepared in two different hyperfine states labeled by a 
spin index $\sigma$. 
Species $b$, which can be a fermion or boson, is subject to a strong harmonic
confining potential.
The bound states for the hyperfine states correspond to the unscreened
spin in the Kondo problem.  In order for the Kondo effect to occur the bound
states need to obey certain conditions.
First, the bound state between the $a$ and $b$ atoms needs to be well occupied,
however, the atoms should not be too tightly 
bound such that second order spin flip processes - important for the Kondo
effect - can occur frequently. This implies a certain regime for the effective
scattering lengths $a_{\sigma}$ bounded by $c_u$ and $c_l$ as shown in
Fig.~\ref{fig:1}~(b). 
Second, the bound states energies should be very similar, since a difference of their
energy will favor a certain polarization of the spin, like a magnetic field,
which suppresses the Kondo effect. 
The bound state energies of atom $a$ with $b$, $E_{b,\sigma}$, generally depend on the
scattering lengths $a_{\sigma}$, $E_{b,\sigma}\simeq -\hbar^2/2 m_r
a_{\sigma}^2$, $m_r=m_am_b/(m_a+m_b)$, and for a general
system those will be very different. The challenge is then to find a scheme in
which the bound states can be simultaneously tuned to an energy which is
roughly equal and in a suitable regime for Kondo physics. 
As we will show, for certain hyperfine states of a system
of ${}^{40}$K and ${}^{23}$Na atoms the effective scattering
lengths intersect in this regime when tuned close to Feshbach
resonances such that Kondo physics is directly possible. 
However, for many other systems this fortuitous meeting of the conditions will
not occur. 
In the Supplementary Material (SM) we show that additional resonances of the
confining potential can also be employed to tune the system into the
Kondo-correlated state \cite{Ols98,MC06}.

\paragraph{Physical setup and effective model -} 
\colortxt{
To formalize these ideas, we first discuss the
atomic scattering problem and then relate the parameters to the
low energy effective model in Eq.~(\ref{HAnd}), which is 
directly connected to the Kondo effect.} Consider for each 
component $\sigma$ the two-particle scattering between species $a$ and
$b$ described by a Hamiltonian of the form,  
\begin{equation}
  H_{\rm
    scat}=\frac{p_b^2}{2m_b}+\frac{1}{2}m_b\omega_{\rm ho}^2r_b^2+\frac{p_a^2}{2m_a}
   +V(\vct r_a-\vct r_b),
\label{eq:Hscat}
\end{equation}
where $\vct p_\alpha$, $\vct r_\alpha$ are momenta and positions of the
particles, $V(\vct r)$ is the interspecies potential and $\omega_{\rm ho}$ a
scale for the harmonic confinement.
A corresponding length scale is the harmonic oscillator length
$a_{\rm ho}=\sqrt{\hbar/m_b\omega_{\rm ho}}$. 
The low energy form of the effective s-wave scattering amplitude
$f_{\sigma}(k)$ in terms of $a_{\sigma}$ and the effective radius
$r_{e,\sigma}$ is,  
\begin{equation}
  f_{\sigma}(k)=\frac{1}{-\frac{1}{a_{\sigma}}+r_{e,\sigma}\frac{k^2}{2}-ik}.
\label{eq:fslowkgen}
\end{equation}
Without the harmonic confinement ($\omega_{\rm ho}=0$) the scattering problem for
each $\sigma$ is characterized by the bare s-wave scattering length
$a_{0,\sigma}$. In presence of the harmonic trap the effective parameters $a_{\sigma}$ and
$r_{e,\sigma}$ can be calculated depending on the bare scattering length
$a_{0,\sigma}$ \cite{MC06,NT10}. One finds in the Born approximation,
\begin{equation}
  \label{eq:Bornapprox}
a_{\sigma}=\frac{m_a}{m_r}a_{0,\sigma},\;\;
r_{e,\sigma}=-\frac{m_r}{m_a}\frac{a_{\rm ho}^2}{a_{0,\sigma}}.  
\end{equation}
To tune the bare scattering lengths $a_{0,\sigma}$ by a magnetic field $B$, we
assume that there is a Feshbach resonance, 
\begin{equation}
  \label{eq:scattfeshbach}
  a_{0,\sigma}(B)=a_{{\rm bg}}\Big(1-\frac{\Delta B_{0,\sigma}}{B-B_{0,\sigma}}\Big),
\end{equation}
where $a_{{\rm bg}}$ is the background scattering length, $\Delta B_{0,\sigma}$ the
width and $B_{0,\sigma}$ the position of the resonance.

We describe the low energy physics of the system by an Anderson impurity model
(AIM) \cite{And61} of the form,    
\begin{eqnarray}
  H&=&\sum_{\vk,\sigma}\epsilon_{\vk} c_{\vk,\sigma}^{\dagger}c_{\vk,\sigma} +
  \sum_{\sigma}\epsilon_{b,\sigma} 
  c_{b,\sigma}^{\dagger}c_{b,\sigma}+U n_{b,\uparrow}n_{b,\downarrow}
  \nonumber \\
&&  + \sum_{\vk,\sigma} V_{\vk,\sigma}c_{\vk,\sigma}^{\dagger}c_{b,\sigma} +\hc \; .
\label{HAnd}
\end{eqnarray}
Here, $c_{\vk,\sigma}^{\dagger}$ creates an itinerant fermion with momentum
$\vk$ and spin projection $\sigma$, and 
$c_{b,\sigma}^{\dagger}$ a bound state with energy $\epsilon_{b,\sigma}$. The
states are mixed by the hybridization $V_{\vk,\sigma}$. 
\colortxt{Three-particle bound states are assumed to be highly unstable
  due to rapid decay into deep two-body bound states \cite{BH06,Kea06}. Under
  such conditions it has been shown \cite{SBLVDGCRD08} that the system
  corresponds to $U$$\to$$ \infty$, where the occupation of those states is
  suppressed. Particle loss is inhibited in such situations
  \cite{SBLVDGCRD08}.}  
The effect of the \colortxt{shallow} trapping potential on the atoms with mass $m_a$ is
neglected, and hence the dispersion is  $\epsilon_{\vk}=\hbar^2 \vk^2/2m_r$
 \footnote{\colortxt{This is justified by the local density approximation,
     which is usually a good description for trapped systems.}}.
We focus on the case of three spatial dimensions, where the corresponding density
of states (DOS) per spin is $\rho_0(\epsilon)=c_3\sqrt{\epsilon}$, with 
$c_3=V_0\kF^3/(4\pi^2 \epsF^{3/2})$, and $\epsilon_{\rm
  F}=\frac{\hbar^2}{2m_r}\left(3\pi^2n\right)^{2/3}$. Here, $n=N_a/V_0$, where
$V_0$ is the volume of the system and $N_a$ the number of particles of species
$a$.

\paragraph{Relation of AIM parameters to scattering parameters -}
\colortxt{The scattering amplitude is related to the $T$-matrix by
\begin{equation}
  f_{\sigma}(\vk,\vk)=-V_0\frac{m_r}{2\pi\hbar^2} T^{\sigma}_{\vk,\vk}.
\label{eq:fTid}
\end{equation}
We can express the right hand side in terms of scattering properties of the AIM,
$T^{{\rm A},\sigma}_{\vk,\vk}(\omega=\epsilon_{\vk})$, and from this determine the 
model parameters in Eq.~(\ref{HAnd}).
The $T$-matrix of the AIM is given by \cite{hewson},}
\begin{equation}
  T^{{\rm A},\sigma}_{\vk,\vk'}(\omega)=V_{\vk,\sigma}^*G_{b,\sigma}(\omega)V_{\vk',\sigma},
\label{eq:TmatAIM}
\end{equation}
where $G_{b,\sigma}(\omega)$ is the retarded bound state Green's function, $\eta\to0$,
\begin{equation}
  G_{b,\sigma}(\omega)^{-1}={\omega+i\eta
    -\epsilon_{b,\sigma}-K_{\sigma}(\omega)-\Sigma_{\sigma}(\omega)}.
\label{eq:Gb}
\end{equation}
Generally, impurity properties only depend on the integrated hybridization
term $  K_{\sigma}(\omega)=\sum_{\vk}\frac{|V_{\vk,\sigma}|^2}{\omega+i\eta
    -\xi_{\vk}}$, where we defined $\xi_{\vk}=\epsilon_{\vk}-\mu$.
 
\colortxt{A full solution of the scattering problem in Eq.~(\ref{eq:Hscat}) yields
$f_{\sigma}$ \cite{MC06} such that the AIM parameters can be determined
numerically via Eq.~(\ref{eq:fTid}). Here we take a
simplified form, $V_{\vk,\sigma}=V_{\sigma}$ , which is real and constant for
$\epsilon_{\vk}< \Lambda_{v,\sigma}$ and zero otherwise. This allows us to derive explicit
analytical expressions. For generic forms  of free states and an s~-wave bound
state, one sees that the overlap integrals 
$V_{\vk,\sigma}$ vanish when the wave length $1/k$ becomes much shorter than
the typical extension of the bound state $a_{\sigma}$. From this follows that
a reasonable assumption is $\Lambda_{v,\sigma}=\frac{\alpha^2}{(\kF
  a_{\sigma})^2}\epsF$, with $\alpha\simeq 1$, used in the following.
From Eq.~(\ref{eq:fTid}) we find then for small $k$ with
Eq.~(\ref{eq:fslowkgen}) and Eq.~(\ref{eq:TmatAIM}) with $\mu\to 0$,}
\begin{equation}
\frac{V_{\sigma}^2}{\epsF^2}=\frac{\frac{8\pi}{V_0\kF^3}}
{\frac{2}{\pi}|\kF a_{\sigma}| - \kF r_{e,\sigma}}, 
\label{eq:Vsig}
\end{equation}
such that $V_{\sigma}^2\sim 1/N_a$ and
\begin{equation}
  \frac{\epsilon_{b,\sigma}}{\epsF}=
 \frac{V_0\kF^3}{4\pi}\Big(\frac{2}{\pi|\kF a_{\sigma}|}-\frac{1}{\kF
     a_{\sigma}}\Big)\frac{V_{\sigma}^2}{\epsF^2}
=2\frac{\frac{2}{\pi|\kF a_{\sigma}|}-\frac{1}{\kF
     a_{\sigma}}}{\frac{2}{\pi}|\kF a_{\sigma}| - \kF
   r_{e,\sigma}}. 
\label{eq:ebsig}
\end{equation}
Useful quantities are the hybridization parameter,
$\Gamma_{\sigma}=\pi\rho_0(\epsF)V_{\sigma}^2$,
which is independent of the volume of the system, and the \colortxt{important} ratio  
\begin{equation}
  \frac{-\epsilon_{b,\sigma}}{\pi\Gamma_{\sigma}}=\frac{1}{\pi}
\left(\frac{1}{\kF  a_{\sigma}}-\frac{2}{\pi|\kF  a_{\sigma}|}\right),
\label{eq:Kondoratio}
\end{equation}
which only depends on $\kF a_{\sigma}$.
We can see how the AIM model parameters depend on $a_{\sigma}$ and
$r_{e,\sigma}$, for instance, $-\epsilon_{b\sigma}$ increases with
$1/a_{\sigma}$ for $a_{\sigma}>0$. As discussed, $a_{\sigma}$ and
$r_{e,\sigma}$ depend on $a_{0,\sigma}$ and thus on the magnetic field
$B$ and the trapping frequency $\omega_{\rm ho}$, and this allows to tune the
model parameters in Eq.~(\ref{HAnd}). \colortxt{We see that in general 
both $h=(\epsilon_{b,\uparrow}-\epsilon_{b,\downarrow})/2$, which acts as a
local magnetic field, and $\Delta \Gamma
=(\Gamma_{\uparrow}-\Gamma_{\downarrow})/2$ can be non-zero.} For studies of
the AIM the latter is unusual, but it has been discussed in situations of
quantum dots coupled to partly polarized leads \cite{Mea03a,Mea03b,CSR04}. 
As discussed in the SM, this
implies that in general due to second order processes in the hopping an
effective local magnetic field $h_{\rm eff}$ is generated. 
There, also the
mapping of the general AIM in Eq.~(\ref{HAnd}) to an anisotropic Kondo model 
with spin-spin couplings $J_{\perp}\neq J_{z}$ and the derivation of the Kondo
scale $T_{\rm K}$ from scaling equations are explained.

\paragraph{Tuning to the Kondo state -}
We now discuss appropriate parameter regimes to observe the Kondo-correlated
state in more detail. 
These are meant as guidelines and not strict boundaries. 
The first condition (I) is to have small fluctuations of the occupation of the
bound state, which can be expected if
$-\epsilon_{b,\sigma}/(\pi\Gamma_{\sigma})>c_1$. 
A naive estimate is $c_1\sim 1/2$, however, 
smaller values, $c_1\sim 0.25$, also work well as shown later.   
The second condition (II) is to have the Kondo scale in a regime where it can be
observed experimentally, i.e., the experimental temperature $T_{\rm exp}\sim
T_{\rm K}$. We assume $T_{\rm  exp}=\alpha_T \epsF$, where $\alpha_T\sim 0.01$
can be achieved. To achieve this the bound state must not lie too deep. 
More formally, $T_{\rm K}$ depends exponentially on the Kondo coupling $J_z$
and the asymmetry $x=J_{\perp}/J_z<1$. It follows that $J_z$ must not be too
small and the asymmetry should not be too large. These couplings depend on the
AIM parameters $\epsilon_{b,\sigma}, 
\Gamma_{\sigma}$ (see SM) and we can obtain a condition of
the form $-1/2\sum_{\sigma} \epsilon_{b,\sigma}/(\pi\Gamma_{\sigma})<c_2$ with
a numerical estimate  $c_2\approx 0.6$.
Hence, together with (I) we define an interval for
values of  $-\epsilon_{b,\sigma}/(\pi\Gamma_{\sigma})$, and with the help of
Eq.~(\ref{eq:Kondoratio}) we can state it as a condition for $\kF a_{\sigma}$, 
\begin{equation}
  \label{eq:kondocond1}
  c_l<\kF a_{\sigma}<c_u,
\end{equation}
where $c_l=\frac{\pi-2}{\pi^2c_2}$, $c_u=\frac{\pi-2}{\pi^2c_1}$ are lower/upper
boundaries [see Fig.~\ref{fig:1}~(b)] .
For $\Delta \Gamma\neq 0$, an effective magnetic field is
generated, which suppresses Kondo correlations. It is possible to offset the
effective field with a local magnetic field $h$, and thus, we define a third
condition (III), 
$\Delta \Gamma=\alpha_h h.$
For a symmetric DOS, $\alpha_h\simeq 0.2$ was found numerically \cite{Mea03b},
however, for a general DOS and different cut-offs this may vary (see
SM). 

To be able to tune to the Kondo-correlated state, it is necessary
to have a system, where the Feshbach 
resonances for $\ket{\uparrow}$ and $\ket{\downarrow}$ with the impurity
atom have some overlap. 
Using the form in Eq.~(\ref{eq:scattfeshbach}) for $a_{0,\sigma}$, $\Delta B_{0,\uparrow}$ and
$\Delta B_{0,\downarrow}$ need to have the same sign and $|B_{0,\sigma}+\Delta
B_{0,\sigma}|>|B_{0,-\sigma}|$. 
Given the bare scattering lengths $a_{0,\sigma}$ the effective scattering
length including the harmonic potential can be calculated. For simplicity we
use the Born approximation in Eq.~(\ref{eq:Bornapprox}) in the following,
which was shown to give  reasonably accurate results \cite{MC06}. Hence, the effective scattering
lengths can be tuned close to the Feshbach resonance and it is possible that
at the intersection, $a_{\uparrow}=a_{\downarrow}$, Eq.~(\ref{eq:kondocond1})
is satisfied. This is illustrated in Fig.~\ref{fig:1} (b).
As $a_{0,\uparrow}=a_{0,\downarrow}$, Eq.~(\ref{eq:Bornapprox}) implies
$r_{e,\uparrow}=r_{e,\downarrow}$, such that \colortxt{$h$$=$$\Delta \Gamma$$ =$$ 0$} and
automatically conditions (I-III) are satisfied. Note 
that if $a_{\uparrow}=a_{\downarrow}$, but $\kF a_{\sigma}>c_u$, we can
satisfy Eq.~(\ref{eq:kondocond1}) by reducing the density $n$ and thus $\kF$.

\begin{figure}[!t]
\centering
\subfigure[]{\includegraphics[width=0.49\columnwidth]{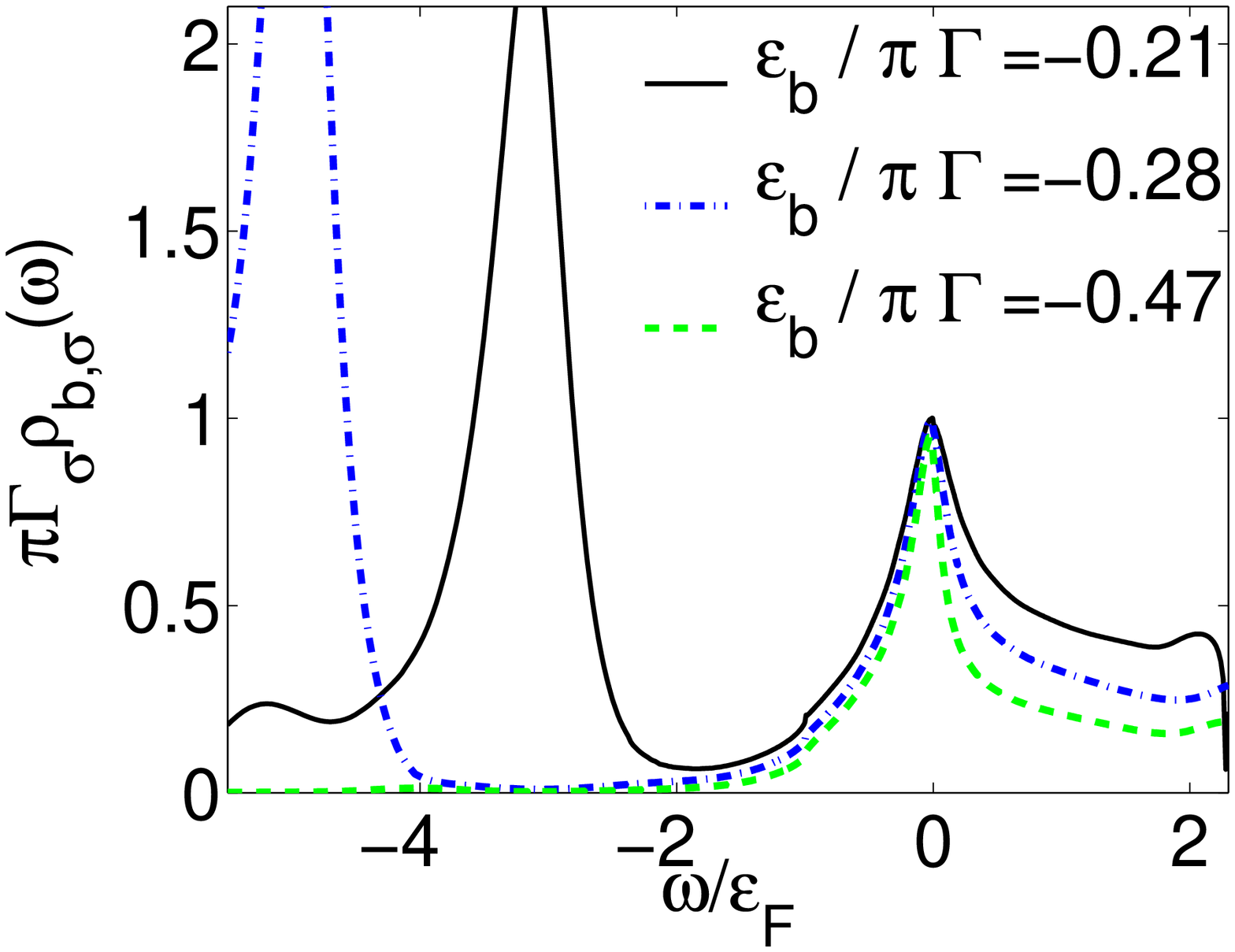}}
\subfigure[]{\includegraphics[width=0.49\columnwidth]{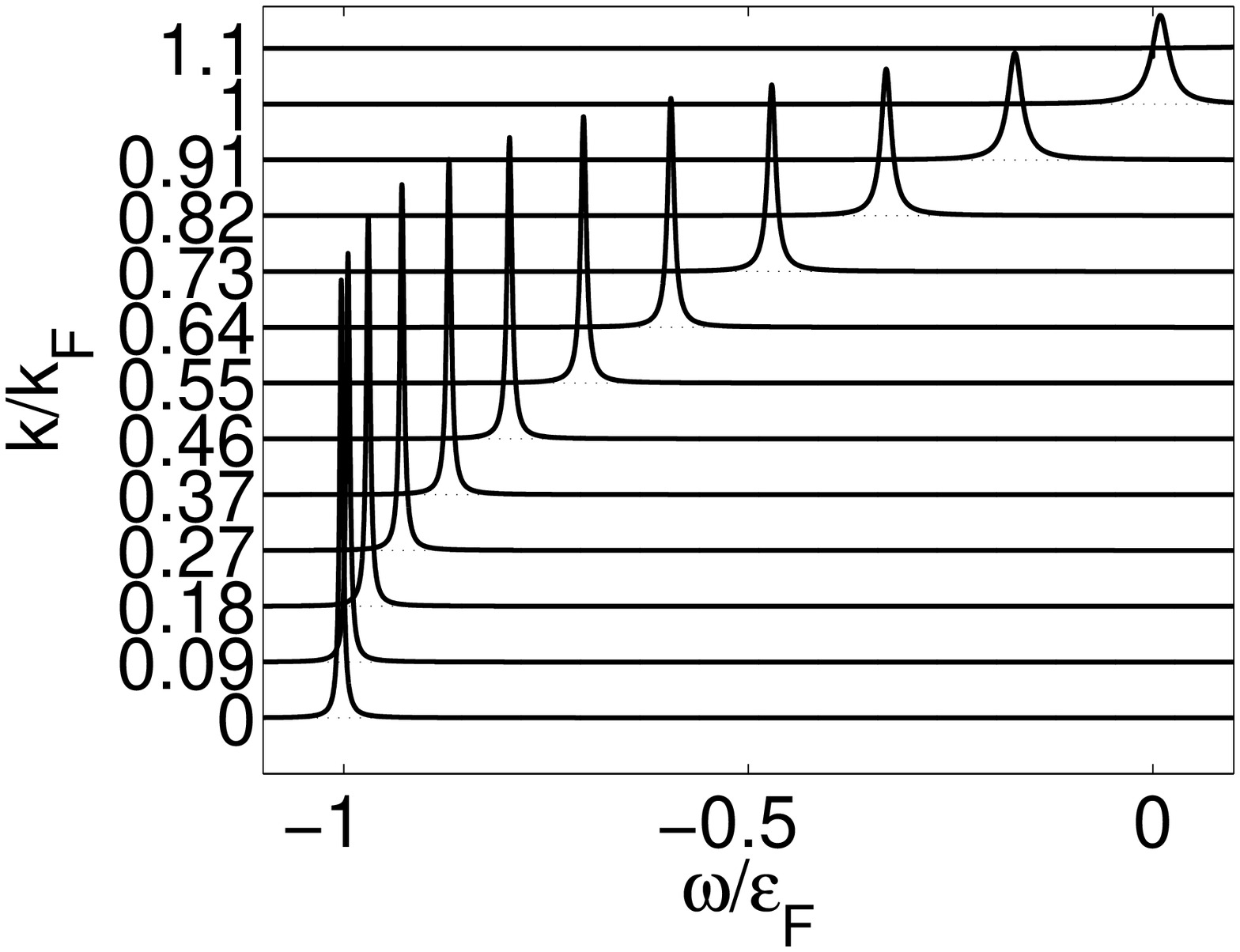}}
\vspace*{-0.5cm}
\caption{(Color online) (a) Spectral function $\rho_{b,\sigma}(\omega)$ for
  different values of $\epsilon_{b,\sigma}/\pi\Gamma$ obtained for  
  $\kF=1,0.75,0.45\kF^0$. The corresponding values are $\kF
  a_{\sigma}=0.55,0.41,0.25$ and the complete set of AIM parameters is given
  in Table~I in the SM. The magnetic field
  $B$ has been tuned such that $a_{\uparrow}=a_{\downarrow}$. (b) Momentum
  resolved spectral function $\rho_{\vk,\sigma}(\omega)$ as calculated from
  Eq.~(\ref{eq:Gbk}) for the case $\epsilon_b/\pi\Gamma=-0.21$ with $n_{\rm
    imp}$ as discussed below   Eq.~(\ref{eq:Gbk}).}
\label{fig:spec_difkF}
\vspace*{-0.5cm}
\end{figure}

\paragraph{Experimental system and probe-}
For the experimental realization of our proposal 
we focus on a system of ${}^{40}$K with 
$\ket{\!\!\downarrow}$~$\equiv$~$\ket{9/2,-7/2}$ and
${\ket{\!\uparrow}}$~$\equiv$~$\ket{9/2,-5/2}$ hyperfine states and ${}^{23}$Na in the 
hyperfine state $\ket{1,1}$. In this system recently molecular
states were successfully produced using interspecies Feshbach resonances
\cite{WPAWZ12}. For our purpose, the Feshbach resonances with $B_{0,\uparrow}=106.9$G,
$B_{0,\downarrow}=108.6$G, $\Delta B_{0,\uparrow}=-1.8$G, and $\Delta
B_{0,\downarrow}=-6.6$G \cite{Pea12} are suitable. The background scattering
length is $a_{\rm bg}=-690 a_{\rm B}$, where the Bohr radius is $a_{\rm B}=0.53\cdot 10^{-10}$m. 
In the following we use common values in ultracold gas experiments,
$n\sim 10^{18}$m${}^{-3}$, $\kF^0=[3\pi^2 n]^{1/3}=1.6\cdot 10^{-4}/a_{\rm B}$. 
A typical trapping frequency is $\omega_{\rm ho}\sim 100$kHz \cite{BDZ08},
such that for ${}^{23}$Na we have $a_{\rm ho}=3.12 \cdot 10^3a_{\rm
  B}$.
One has $a_{0,\uparrow}=a_{0,\downarrow}=-1.82 a_{\rm bg}$ for 
\begin{equation}
  B_s=\frac{B_{0,\downarrow}\Delta B_{0,\uparrow}-B_{0,\uparrow}\Delta
    B_{0,\downarrow}}{\Delta B_{0,\uparrow}- \Delta B_{0,\downarrow}}=106.26
  {\rm G}.
\label{eq:Bs}
\end{equation}
With Eq.~(\ref{eq:Bornapprox}), this implies $\kF^0 a_{\sigma}=0.55$.
Employing the numerical renormalization group (NRG) \cite{Wil75,BCP08}, we
calculated the low temperature bound state spectral function, 
$\rho_{b,\sigma}(\omega)$, for parameters corresponding to the 
K-Na-system for the situation where $B=B_s$ in Eq.~(\ref{eq:Bs}) and we vary
$\kF/\kF^0$, where $\kF^0$ is as above. The energy scale is set by $\epsF=1$,
and we measure all energies from $\epsF$. The numerical values for the AIM
parameters are given in Table~I in the SM. The result is shown in 
Fig.~\ref{fig:spec_difkF} (a). 
The bound state peaks lie between $-8\epsF$ (not shown) and $-2\epsF$. 
Due to interaction effects the peaks are broadened, even though they lie
outside the continuum, however, the width might be overestimated in the NRG calculation.   
At $\omega=0$ we see a clear peak, the Kondo or Abrikosov-Suhl resonance.  We can
extract the half width of the Kondo resonance $\Delta_{\rm K}\sim 0.1 \epsF$, which
is indicative of the Kondo scale suitable for experimental observations. 

\noindent
We will now describe experimental signatures to detect the Kondo-correlated state. 
The retarded Green's function of itinerant states in presence of $n_{\rm imp}$ impurities
is given by \cite{hewson},
\begin{equation}
  G_{\vk,\sigma}(\omega)^{-1}={\omega+i\eta-\xi_{\vk}-n_{\rm imp} V_{\sigma}^2
    G_{b,\sigma}(\omega)},
\label{eq:Gbk}
\end{equation}
and $\rho_{\vk,\sigma}(\omega)=-\frac1{\pi}\Imag G_{\vk,\sigma}(\omega)$,
$\eta\to 0$, is the momentum resolved spectral function.
In ultracold gas experiments, $\rho_{\vk,\sigma}(\omega)$ can be measured directly by momentum resolved
photo-emission spectroscopy \cite{PDGDSC07,SGJ08,FFVKK11}. 
For an impurity concentration $n_{\rm imp}/N_a\approx 0.03$ and parameters
corresponding to $\epsilon_b/\pi\Gamma=-0.21$ in Fig.~\ref{fig:spec_difkF}~(a),
we show $\rho_{\vk,\sigma}(\omega)$  in Fig.~\ref{fig:spec_difkF}~(b).
The change of the height and width of the peaks when approaching $\kF$ can be
easily understood as due to the coupling of the itinerant states to
the Kondo resonance. In Eq.~(\ref{eq:Gbk}) the imaginary part of $G_{b,\sigma}(\omega)$,
which is proportional to the spectral function $\rho_{b,\sigma}(\omega)$  leads
to a broadening of the spectral function, which is most pronounced close to
$\epsilon_{\vk}=\epsF$, where the Kondo resonance lies. This is a striking
feature opposite to usual scattering mechanisms which increase when
moving away from the Fermi energy. In fact, we can
interpret this is as a self-energy term, $\Sigma(\omega)=n_{\rm imp} V_{\sigma}^2
G_{b,\sigma}(\omega)$, and this can be extracted
\footnote{\colortxt{The energy resolution required is set by the peak width
    in Fig.~\ref{fig:spec_difkF}~(b). For $n_{\rm imp}/N_a \sim 0.1$ the
    width at $k=\kF$ is $\Delta/\epsF \sim 0.09$.}} by well established 
techniques developed in angle-resolved photo emission spectroscopy
\cite{Vea99,DSH03}.
Hence, this procedure gives access to the spectral function in
Fig.~\ref{fig:spec_difkF}~(a) and as such is a direct signature of the Kondo
peak.
\colortxt{More conventional radio frequency (rf) spectroscopy can also provide explicit
signatures of the Kondo-correlated state. As discussed comprehensively in the
SM the signal shows characteristic broadened peaks, which are shifted from the
ones for a system without the Kondo effect.}
Also spectroscopy of the species $b$ impurity atom, switching from
an uncorrelated state to a Kondo state, would be very
interesting. Characteristic power law tails can be observed as discussed in
detail in Refs.~\cite{Lea11,Kea12}.

\paragraph{Addressing open questions -}
 
We discuss now a number of interesting questions, which can be addressed once
the Kondo-correlated state has been realized. It has been long argued \cite{hewson}
that the magnetically screened impurity should be surrounded by a screening
cloud with the spatial extension of order $\xi_{\rm K}=\frac{\hbar v_{\rm F}}{k_{\rm
    B}T_{\rm  K}}$. However, experimentally it has not been possible to
provide firm evidence for the Kondo cloud, such that its existence is unclear.
A typical quantity to show Kondo cloud features is the decay of the equal time spin
correlation function $\expval{S_c(\vct r)s_{\rm imp}}{}$, which is difficult
to access in condensed matter systems. In contrast, in the proposed ultracold gas system this
correlation function \colortxt {should become} accessible by spectroscopic
tools \cite{KKGBLD13}, such that
$\xi_{\rm K}$ could be determined and this fundamental question of Kondo
physics be settled. Since our setup allows to switch the impurity on and off
by optically changing the hyperfine state of the $b$ species,
it would also be curious to analyze the time scale on which the Kondo cloud
builds up. 
The proposed setup has also great potential to shed light on a number of
intriguing issues for the Kondo lattice including the observation of a
fractionalized Fermi liquid (FL${}^*$) 
phase \cite{SSV03} and the occurrence of superconductivity close to a quantum
phase transitions.

\paragraph{Conclusions -}
We have demonstrated how to realize a Kondo-correlated
state for a mixture of ultracold atoms. The proposed setup, different
from a previous proposal based on alkaline-earth atoms \cite{Gea10}, ones
related to the spin-Boson model \cite{RFZVZ05,OSL08} and a
bosonic form \cite{FDS04}, can be realized with experimental techniques
currently available. In general, this allows one to analyze
field dependent and anisotropic Kondo physics.  We proposed the recently
studied Na-K-mixture as a suitable system and computed the rf response in a 
regime well accessible by experiments. There are numerous possible
extension of our work including the study of non-equilibrium Kondo physics,
Kondo lattice systems and signatures of quantum criticality. 
We point out that geometrical resonances which we discuss in the 
supplementary material can also appear in $l >0$ angular momentum 
channels even for purely s-wave scattering between localized and 
itinerant atoms (see e.g. \cite{MC06,NT10}). This may open 
interesting possibilities for realizing multichannel Kondo physics.

\paragraph{Acknowledgments -} We wish to thank S. Blatt, S. Gopalakrishnan,
A.C. Hewson, M. Koehl, Y. Oreg, M. Punk, S. Will, A. Tsvelik, G. Zarand, F. Zhou, and
M. Zwierlein for helpful discussions. We would like to thank Martin Zwierlein
for pointing out to us the possibility of using Feshbach bound states on impurity
atoms to realize the Kondo effect in cold atomic systems and for a collaboration on
related projects \cite{VPZD11}. JB acknowledges financial support from the DFG
through grant number BA 4371/1-1. We also acknowledge support from Harvard-MIT
CUA, DARPA OLE program, AFOSR Quantum Simulation MURI, AFOSR MURI on Ultracold
Molecules, the ARO-MURI on Atomtronics, ARO MURI Quism program and the EU ERC
Ferlodim.

\newpage

\begin{appendix}
\section{S1 Hybridization function}
\begin{figure}[!htbp]
\centering
\includegraphics[width=0.65\columnwidth]{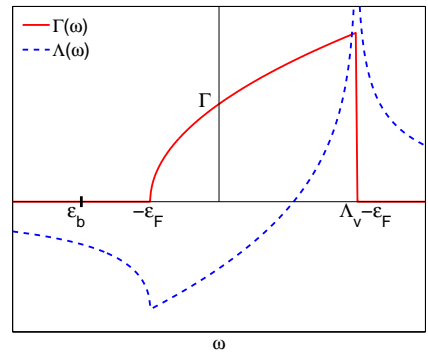}
\caption{(Color online) Schematic plot of the hybridization function
  $\Gamma(\omega)$ and $\Lambda(\omega)$ indicating the 
  boundaries of the continuum states $-\epsF$ and upper cutoff $\Lambda_v$ and
  the energy of the bare bound state $\epsilon_b$, generally below the
  continuum. At the sharp positive edge $\Lambda(\omega)$ has a logarithmic
  singularity. This can lead to bound states outside the continuum
  \cite{BVZ06}. The sharp cut-off can be softened without changing any of the
  results qualitatively. For simplicity we have omitted the index $\sigma$.}
\label{fig:gamma}
\end{figure}

The  hybridization function was defined as,
\begin{equation}
  K_{\sigma}(\omega)=\sum_{\vk}\frac{|V_{\vk,\sigma}|^2}{\omega+i\eta -\epsilon_{\vk}+\mu}.
\end{equation}
We write $K_{\sigma}(\omega)=\Lambda_{\sigma}(\omega)-i\Gamma_{\sigma}(\omega)$.  
If we assume that $V_{\vk,\sigma}=V_{\sigma}$ is constant and is cut off
by $\Lambda_{v,\sigma}$, then the imaginary part becomes, 
\begin{equation}
  \Gamma_{\sigma}(\omega)=-\pi|V_{\omega,\sigma}|^2\rho_0(\omega+\mu)\theta(\omega+\mu)
  \theta(\Lambda_{v,\sigma}-\mu-\omega). 
\label{eq:Gamma}
\end{equation}
The real part can also be evaluated in terms of principle value
integrals and we find for $\omega<-\mu$,  
\begin{equation}
  \Lambda^{<}_{\sigma}(\omega)=-2|V_{\sigma}|^2c_3\Big[\sqrt{\Lambda_{v,\sigma}}-\sqrt{-\omega-\mu}
  \arctan\Big(\frac{\Lambda_{v,\sigma}}{-\omega-\mu}\Big)\Big]. 
\label{eq:Lambda}
\end{equation}
and for  $\omega>-\mu$,
\begin{eqnarray*}
   \Lambda^{>}_{\sigma}(\omega)&=&-2|V_{\sigma}|^2c_3\Big[\sqrt{\Lambda_{v,\sigma}}\\
&& -\frac12\sqrt{\omega+\mu}
  \log\Big(\frac{\sqrt{\Lambda_{v,\sigma}}+\sqrt{\omega+\mu}}
{\sqrt{\Lambda_{v,\sigma}}-\sqrt{\omega+\mu}}\Big)\Big].  
\end{eqnarray*}
The function is continuous at $\omega\to -\mu$ but not smooth. A schematic
plot with $\mu=\epsF$ is shown in Fig.~\ref{fig:gamma}.

\section{S2 Kondo physics with additional resonances}
In the main text it was discussed how the system of ${}^{40}$K and ${}^{23}$Na
atom can be tuned into a Kondo-correlated state employing two Feshbach
resonances. It is fortunate in this situation that the effective scattering
lengths intersect in a suitable regime as show in Fig.~1(b). However, for
other systems this will not generally be the case.  
For instance, two hyperfine states of ${}^{6}$Li have a number of Feshbach
resonances with ${}^{133}$Cs between 800-900G with promising features
\cite{Tea13}, but the intersection does not directly lie in a suitable regime.
Here we show that it can nevertheless be possible to tune the system into a
Kondo-correlated state with the help of confinement induced additional
resonances \cite{Ols98,MC06}. 

To see how the confining potential can help to align the bound state energies
in the right regime, we consider
for each component $\sigma$ the two-particle scattering problem between species $a$ and
$b$. Without the harmonic confinement ($\omega_{\rm ho}=0$) the scattering problem for
each $\sigma$ is characterized by the bare s-wave scattering length
$a_{0,\sigma}$. For $a_{0,\sigma}>0$, the effective scattering length $a_{\sigma}$ is found
to have many sharp resonances \cite{MC06}, which can be understood as follows.
A molecular bound state, where only $b$ feels the confinement has the harmonic
oscillator energy $E_n=(2n+\frac32)\sqrt{\frac{m_b}{M}}\hbar\omega_{\rm ho}$, with $M=m_a+m_b$.
This energy reduced by the binding energy $E_{b,\sigma}$ can become resonant with the ground
state energy of oscillator and atom, $E_n+E_{b,\sigma}= 3/2\hbar \omega_{\rm
  ho}$, which leads to the resonance for the effective scattering length $a_{\sigma}$.
From this we obtain the condition \cite{MC06},   
\begin{equation}
  a_{0,\sigma}^{\rm res}(n)=\frac{a_{\rm
      ho}}{\sqrt{\frac{2m_r}{m_b}}\sqrt{(2n+\frac
      32)\sqrt{\frac{m_b}{M}}-\frac 32}} 
\label{eq:horescond}
\end{equation}
where $n=1,2,\ldots $. Hence, in a situation where one scattering length is in
suitable regime the second one can  be tuned there by one of those additional
resonances.

\begin{figure}[!htbp]
\subfigure[]{\includegraphics[width=0.51\columnwidth]{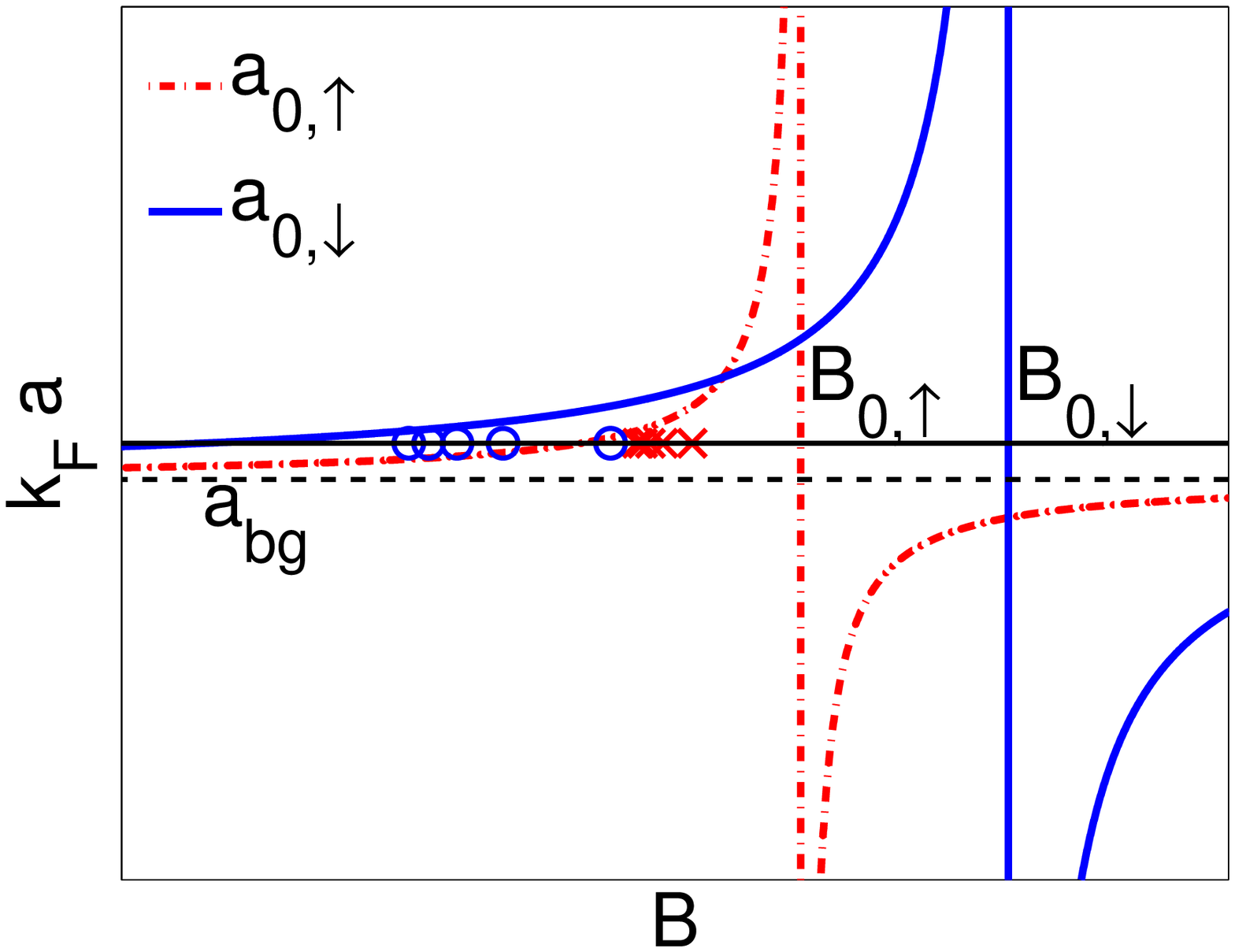}}
\subfigure[]{\includegraphics[width=0.48\columnwidth]{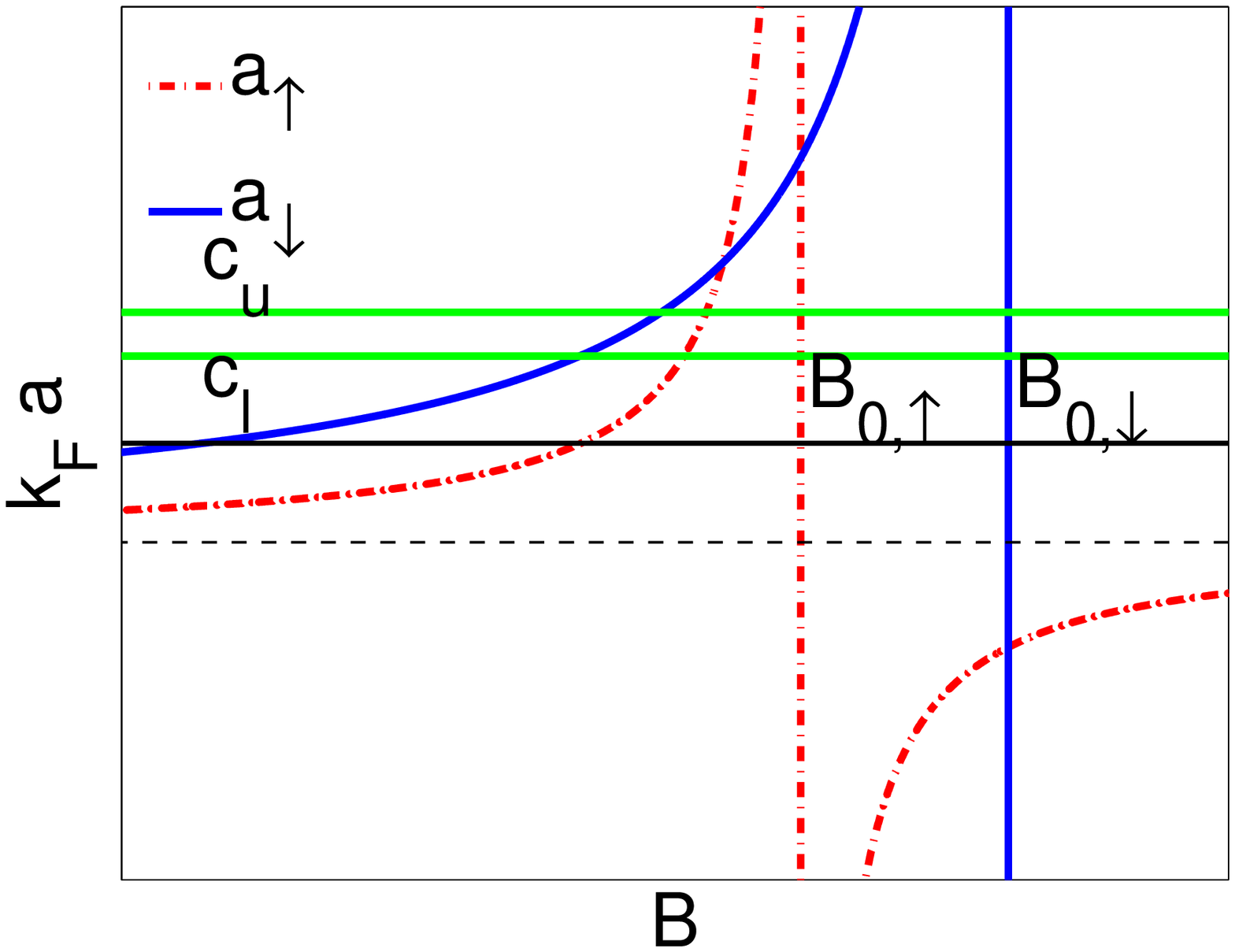}}
\hspace*{0.1cm}
\subfigure[]{\includegraphics[width=0.48\columnwidth]{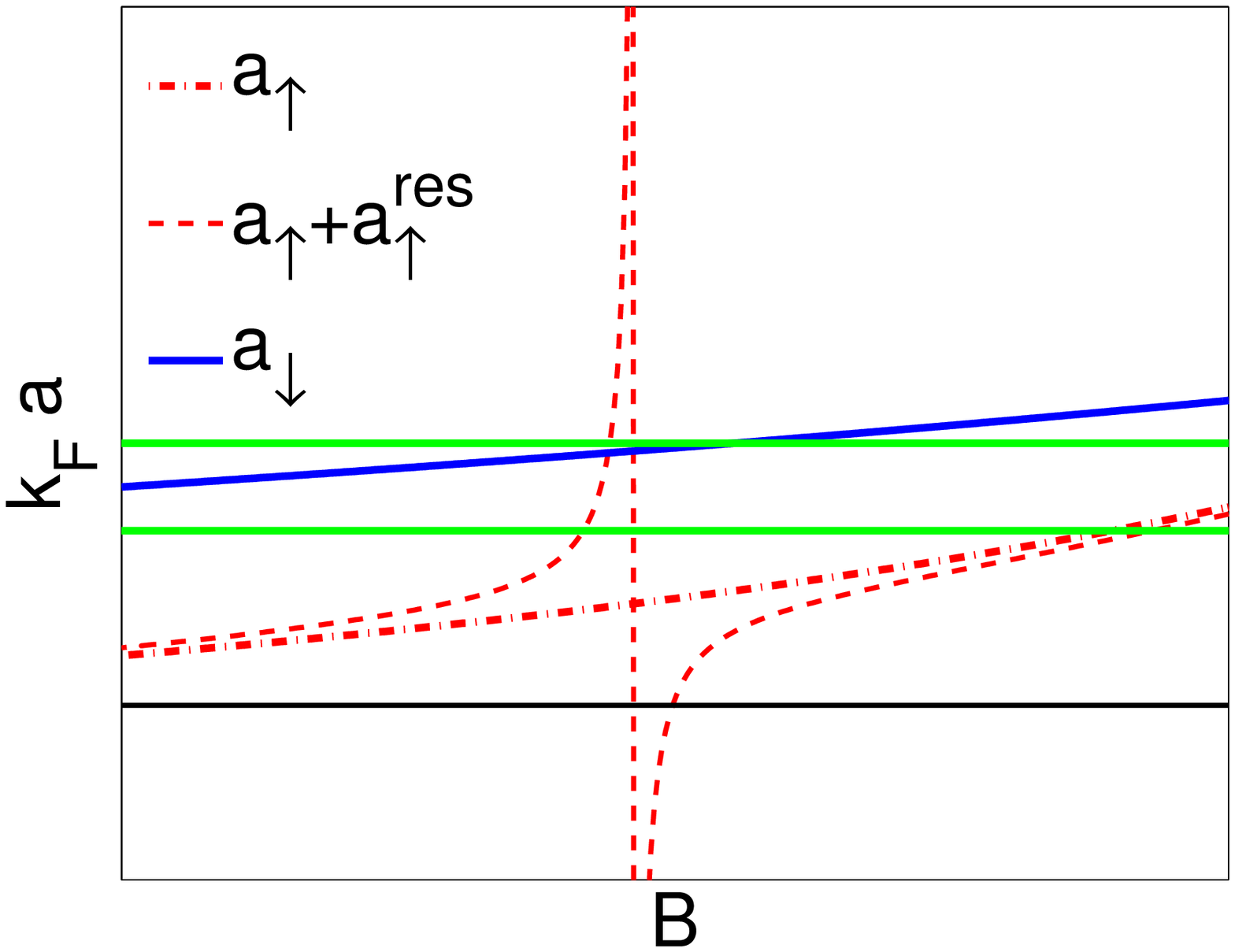}}
\vspace*{-0.3cm}
\caption{(Color online) (a) Schematic plot of the bare scattering length
  $a_{0,\sigma}$ for two  interspecies Feshbach resonances, with
  $B_{0,\uparrow}<B_{0,\downarrow}$, and $|\Delta B_{0,\uparrow}|<|\Delta
  B_{0,\downarrow}|$, $\Delta B_{0,\sigma}<0$, $a_{\rm bg}<0$. The circles
  (crosses) indicate the first 5 magnetic fields where the resonance
  condition in Eq.~(\ref{eq:horescond}) is satisfied for $a_{0,\downarrow}$
  ($a_{0,\uparrow}$). (b) Schematic plot of the effective scattering length
  $a_{\sigma}$ for a situation where the condition $a_{\uparrow}=a_{\downarrow}$ is
  not satisfied within the Kondo boundaries $(c_l,c_u)$. (c) Close-up of (b),
  where a resonance has been tuned to $B=B_{\rm K}$.}  
\label{fig:a0Bdep}
\vspace*{-0.4cm}
\end{figure}

In Fig.~\ref{fig:a0Bdep}~(a) we show schematically the bare scattering lengths
$a_{0,\sigma}$ close to Feshbach resonances. We have also indicated
values of the magnetic field where the resonance condition,
Eq.~(\ref{eq:horescond}), is satisfied.

\noindent
Now let us assume that on tuning $B$ we have 
satisfied $c_l <\kF a_{\downarrow}(B_{\rm K}) < c_u$, but
$\kF a_{\uparrow}(B_{\rm K})< c_l$ [see Fig.~\ref{fig:a0Bdep} (b)].
Then one can change $\omega_{\rm ho}$ by the laser power to bring a sharp
resonance in the vicinity, i.e., change $a_{\rm  ho}$ such that $a_0(B_{\rm
  K})\simeq a_0^{res}(n_{\rm K})$ for some $n_{\rm K}$. Then some further fine tuning to $B_{\rm
  K}-\delta B$ will satisfy the condition $a_{\downarrow}(B_{\rm K}-\delta
B)=a_{\uparrow}(B_{\rm K}-\delta B)$, such that conditions (I) and (II) are
satisfied [see Fig.~\ref{fig:a0Bdep} (c)]. 
Since we have used the additional resonance we have
$r_{e,\uparrow}\neq r_{e,\downarrow}$, and thus in general $h,\Delta \Gamma \neq 0$. There
are two different cases to be considered: (i) $|r_{e,\sigma}|\ll a_{\sigma}$,
in this case, $\Delta \Gamma \approx 0$ and $h\approx 0$, such that conditions
(I-III) are satisfied. Note that $r_{e,\sigma}$ can be reduced by decreasing
$a_{\rm ho}$ as seen in Eq.~(3).  (ii) $|r_{e,\sigma}|\sim
a_{\sigma}$ or  $|r_{e,\sigma}| > a_{\sigma}$, in this case we have to tune
$B$ further to satisfy the condition (III), $\Delta \Gamma=\alpha_h h$.
We can focus on the fast variation of $a_{\uparrow}$ close to the resonance, and assume the
other quantities in this regime as constant. Depending on the strength of the asymmetry
$r_{e,\uparrow}/r_{e,\downarrow}$ a solution which still respects
Eq.~(12) can be found.

\section{S3 Properties of the AIM and mapping to the Kondo model}
Here we discuss some properties of the AIM in
Eq.~(5). First, we show how an effective magnetic
field emerges for the case $V_{\uparrow}\neq V_{\downarrow}$.
This is seen directly  from the scaling equations for the Anderson
model \cite{Hal78,hewson}. 
For a general DOS one finds,
\begin{eqnarray*}
 \frac{d h}{d\Lambda}&=&
\frac{1}{2\pi}\Big[\frac{\Gamma_{\uparrow}(\Lambda)}{\Lambda-\epsilon_{b,\uparrow}}
-\frac{\Gamma_{\downarrow}(\Lambda)}{\Lambda-\epsilon_{b,\downarrow}} \\
&& 
+ \frac{\Gamma_{\downarrow}(-\Lambda)}{\Lambda+\epsilon_{b,\downarrow}+U}-
\frac{\Gamma_{\uparrow}(-\Lambda)}{\Lambda+\epsilon_{b,\uparrow}+U}\Big]. 
\end{eqnarray*}
We introduced
$\Gamma_{\sigma}(\Lambda)=\Gamma_{\sigma}\rho_0(\Lambda+\mu)$. When
considering the situation $U\to \infty$ the last two terms can be neglected.
In the special case of a particle-hole symmetric model, $\epsilon_{b}=-U/2$,
$\epsilon_b=(\epsilon_{b,\uparrow}+\epsilon_{b,\downarrow})/2$, with
symmetric and constant DOS, $\rho_0=1/W$, and initially zero field 
$h=0$, no effective field is generated.  We can see this
from 
\begin{equation}
 \frac{d  h}{d\Lambda}=
\frac{\Gamma_{\uparrow}-\Gamma_{\downarrow}}{2\pi}\frac{2\epsilon_b+U}{(\Lambda+\epsilon_{b}+U)(\Lambda-\epsilon_b)}, 
\label{eq:dh}
\end{equation}
where the right hand side vanishes in the particle hole symmetric case. In all
other cases for $\Gamma_{\uparrow}\neq \Gamma_{\downarrow}$ an effective field
$h_{\rm eff}$ is generated as a second order effect in the hybridization. From
Eq.~(\ref{eq:dh}) we can see that the effective field is proportional to
$\Delta\Gamma$, such that the form $\Delta
\Gamma=\alpha_h  h$ to offset the field is appropriate. We see that for
$\Delta\Gamma >0$, $U+2\epsilon_b>0$ reducing $\Lambda$ leads to a
negative $h_{\rm eff}$. Hence, $\alpha_h>0$ in the offset equation, which is
opposite to the convention in Ref.~\cite{Mea03b}. The above argument is valid
if the hybridization cut-offs $\Lambda_{v,\sigma}$ are equal. However, for
$\Lambda_{v,\uparrow}>\Lambda_{v,\downarrow}$, we find that in the regime
$\Lambda \in (\Lambda_{v,\downarrow},\Lambda_{v,\uparrow})$, $h(\Lambda)$
decreases, since $\Gamma_{\downarrow}(\Lambda)=0$. This has to be taken into
account, when the effective field $h_{\rm eff}$ is computed.

To understand the specific Kondo properties it is convenient to map the AIM in
Eq.~(5) to a Kondo model by a Schrieffer-Wolff
transformation \cite{SW66}. The transformed Hamiltonian is 
\begin{equation}
H'= H_0+\frac 12 \comm{S}{H_1}+\dots\; .  
\end{equation}
$H_1$ is the hybridization term and higher order terms contain higher powers in $V_{\vk,\sigma}$. 
The explicit choice of the generator is,
\begin{eqnarray*}
  S&=&\sum_{\vk,{\sigma}}\Big[\Big(\frac{V_{\vk,\sigma}}{\epsilon_{\vk}-\epsilon_{b,\sigma}}
(1-n_{b,-\sigma})\elcre{\vk}{\sigma}\elann{b}{\sigma} - \hc\Big) \\
&&+\Big(\frac{V_{\vk,\sigma}}{\epsilon_{\vk}-\epsilon_{b,\sigma}-U}
n_{b,-\sigma}\elcre{\vk}{\sigma}\elann{b}{\sigma}-\hc\Big)\Big].
\end{eqnarray*}
This includes projection operators for the occupation of the site.
Evaluating the commutator yields the interaction term of the transformed Hamiltonian, 
\begin{eqnarray*}
\HI&=&\sum_{\vk,\vk'}[J_{\vk,\vk',\downarrow,\uparrow}S^+\elcre{\vk}{\downarrow}\elann{\vk'}{\uparrow}+
    J_{\vk,\vk',\uparrow,\downarrow}S^-\elcre{\vk}{\uparrow}\elann{\vk'}{\downarrow}+ \\
&&  S_z J_{\vk,\vk',z}
    (\elcre{\vk}{\uparrow}\elann{\vk'}{\uparrow}-\elcre{\vk}{\downarrow}\elann{\vk'}{\downarrow})]. 
\end{eqnarray*}
Here, $\vct S$ is the effective impurity spin, 
$S_i=\frac12\elcre{b}{{\sigma}}\sigma^{(i)}_{{\sigma}{\sigma}'}\elann{b}{{\sigma}'}$,
$S^{\pm}=S_1\pm i S_2$. We have introduced,
\begin{eqnarray*}
  J_{\vk,\vk',\sigma,\sigma'}&=&\frac{V_{\vk,\sigma}V^*_{\vk',\sigma'}}{2}
\Big(\frac1{\epsilon_{\vk}-\epsilon_{b,\sigma}}+\frac1{\epsilon_{\vk'}-\epsilon_{b,\sigma'}} \\
&&-\frac1{\epsilon_{\vk}-\epsilon_{b,\sigma}-U}-\frac1{\epsilon_{\vk'}-\epsilon_{b,\sigma'}-U}
\Big).
\end{eqnarray*}
Often one uses $\epsilon_{\vk},\epsilon_{\vk'}\simeq  0$  such that the expressions simplify.
Note that $J_{\vk,\vk',\uparrow,\downarrow}=J_{\vk',\vk,\downarrow,\uparrow}$,
so that $\HI$ is hermitian. 
The first two terms are spin-flip
terms of a spin with the conduction band electrons.
We also have an additional term which behaves like an effective magnetic
field, 
\begin{eqnarray}
H_{\rm mag}&=& S_z\sum_{\vk,\vk'}[\Delta J_{\vk,\vk',z}
(\elcre{\vk}{\uparrow}\elann{\vk'}{\uparrow}+\elcre{\vk}{\downarrow}\elann{\vk'}{\downarrow}) \\
&&-(W_{\vk,\vk',\uparrow}-W_{\vk,\vk',\downarrow})].   
\end{eqnarray}
where $\Delta J_{\vk,\vk',z}=\frac12 (J_{\vk,\vk',\uparrow,\uparrow}-
J_{\vk,\vk',\downarrow,\downarrow})$ and $W_{\vk,\vk',\sigma}$ is given by,
\begin{equation}
  W_{\vk,\vk',\sigma}=\frac{V_{\vk,\sigma}V_{\vk',\sigma}^*}{2}
 \left(\frac1{\epsilon_{\vk}-\epsilon_{b,\sigma}}+\frac1{\epsilon_{\vk'}-\epsilon_{b,\sigma}}\right).
\end{equation}
This term vanishes for a symmetric model $\epsilon_b=-U/2$, a symmetric DOS
and $V_{\uparrow}=V_{\downarrow}$. However, in general it is finite and acts
like a local magnetic field as discussed above for the AIM. 
Generally, there are also terms $\sim~\elcre{b}{\sigma}\elcre{b}{-\sigma}$,
which change the impurity occupation by 2. In the regime of single occupancy
they are neglected. 
We also take the Kondo couplings independent of $\vk$. 
We define $J_{\perp}=J_{\uparrow,\downarrow}=J_{\downarrow,\uparrow}$ and
$J_z=(J_{\uparrow,\uparrow}+J_{\downarrow,\downarrow})/2$, 
\begin{eqnarray*}
J_{\perp}=\frac{V_{\uparrow}V_{\downarrow}}{2}\Big(\frac1{-\epsilon_{b,\uparrow}}+
\frac1{-\epsilon_{b,\downarrow}}\Big),\;
J_{z}=\frac{1}{2}\Big(\frac{V_{\uparrow}^2}{-\epsilon_{b,\uparrow}}+\frac{V_{\downarrow}^2}{-\epsilon_{b,\downarrow}}\Big)  .
\end{eqnarray*} 
In general $J_{\perp}\neq J_z$ and
hence we have to deal with an anisotropic Kondo model.

\section{S4 Scaling of the anisotropic Kondo model and $T_{\rm K}$}

We can define the Kondo scale $T_{\rm K}$ where $J_z$ diverges in the scaling
equations \cite{And70,hewson}. In the
isotropic case for constant DOS,  the result is
\begin{equation}
  T_{0,\rm K}= \Lambda_0 \e^{-\frac{1}{2\rho_0 J}}.
\end{equation}
Here, $\Lambda_0$ is a suitable high energy cutoff.
The square root DOS of the three dimensional systems leads to minor
modifications which can be included in a prefactor. 
In the anisotropic case the equations read,
\begin{eqnarray}
  \frac{dJ_z}{d\log(\Lambda)} &=&-[\rho_0(\Lambda+\mu)+\rho_0(-\Lambda+\mu)]J_{\perp}^2, \\
  \frac{dJ_{\perp}}{d\log(\Lambda)}&=&-[\rho_0(\Lambda+\mu)+\rho_0(-\Lambda+\mu)]J_{\perp}J_z.
\end{eqnarray}
Using $J_z^2-J_{\perp}^2=C$, the Kondo scale $T_{\rm K}$ is found to be 
\begin{equation}
  T_{\rm K}= \alpha_{\rm K}\Lambda_0 \e^{-\frac{\gamma}{2\rho_0J_0}},\qquad
  \gamma=\frac{{\rm atanh}(\sqrt{1-x^2})}{\sqrt{1-x^2}},
\label{eq:TKa1}
\end{equation}
for $x=J_{\perp}/J_z<1$ and $\gamma=\frac{{\rm
    atan}(\sqrt{x^2-1})}{\sqrt{x^2-1}}$ for $x>1$.   
The coefficient $\alpha_{\rm K}$ accounts for the varying DOS and possibly
different upper and lower cutoffs.

Having established these relations, we can derive a condition for $J_z$, such
that $T_{\rm K}$ is of the order of the experimental temperature (see main
text).  It can be written in the form $-1/2\sum_{\sigma}
\epsilon_{b,\sigma}/(\pi\Gamma_{\sigma})<c_2$.  
If one assume that the asymmetry is not very large,
that is $x>1/2$ in Eq.~(\ref{eq:TKa1}), 
one can find the estimate $c_2\approx 0.6$.

\section{S4 NRG calculations for bound state spectral functions}
We have done NRG calculations for the AIM in Eq.~(5) and computed the
low temperature bound state spectral function $\rho_{b,\sigma}(\omega)$. The
$\omega$-dependence of $K_{\sigma}(\omega)$ is dealt with as in NRG applications for
dynamical mean field calculations \cite{BCP08}. The first set of
parameter is for the situation, 
where $B=B_s$ in Eq.~(13) and we vary $\kF/\kF^0$, where $\kF^0=1.6\cdot
10^{-4}/a_{\rm B}$. The parameters are calculated as follows: $a_{\sigma}$ and
$r_{e,\sigma}$ are calculated from Eq.~(3),
and the AIM model parameters $\epsilon_{b,\sigma}/\epsF$, and $V_{\sigma}/\epsF$ are computed from
Eq.~(10) and Eq.~(9). The ratio
$\frac{-\epsilon_{b,\sigma}}{\pi\Gamma_{\sigma}}$ is given in
Eq.~(11), the upper cutoff
$\Lambda_{v,\sigma}=\frac{\epsF}{(\kF a_{\sigma})^2}$, and $\epsF=1$ sets the
energy scale in the calculations. We chose $U$ large enough that double
occupancy is strongly suppressed.   
The numerical values for the parameters are given in Table~\ref{table:kfdep}.
The value for $V_{\sigma}$ corresponds to the choice $V_0\kF^3=1000$.

\begin{table}[!htpb]
\begin{tabular}{| c| c| c| c | c|}\hline
  $\kF/\kF^0$ & $\kF a_{\sigma}$ & $\epsilon_{b,\sigma}/\epsF$ &
  $V_{\sigma}/\epsF$ & $-\epsilon_{b,\sigma}/(\pi\Gamma_{\sigma})$ \\ \hline
 \; 1    \;& \; 0.55 \;&\; -1.64 \;&\; 0.177 \;& 0.21 \\ \hline
 \; 0.75 \;& \; 0.41 \;&\; -2.92 \;&\; 0.204 \;& 0.28 \\ \hline
 \; 0.45 \;& \; 0.25 \;&\; -8.11 \;&\; 0.264 \;& 0.47 \\ \hline
\end{tabular}
 \caption{Values for the AIM model parameters, when varying the density.}
  \label{table:kfdep}
\vspace*{0.3cm}
\begin{tabular}{| c| c| c| c | c| c| c | c|}\hline
  $B (G)$ & $\kF a_{\uparrow}$ & $\kF a_{\downarrow}$ & $\epsilon_{b}/\epsF$ &
  $ h/\epsF$ & $V_{\uparrow}/\epsF$ & $V_{\downarrow}/\epsF$ & $\Delta\Gamma/\epsF$ \\ \hline
  106.25 &  0.55 & 0.55  & -1.64 & 0.000  &  0.177 & 0.177 &  0.000 \\ \hline
  106.2  &  0.48 & 0.53  & -1.78 & -0.075 &  0.174 & 0.176 & -0.029 \\ \hline
  106.1  &  0.38 & 0.50  & -1.96 & -0.175 &  0.167 & 0.175 & -0.111  \\ \hline
  106.0  &  0.30 & 0.47  & -2.12 & -0.243 &  0.157 & 0.174 & -0.219  \\ \hline
\end{tabular}
 \caption{Values for the AIM model parameters, when varying the magnetic field
   $B$.}
  \label{table:Bdep}
\begin{tabular}{| c| c| c | c| c| c | c|}\hline
  $\kF a_{\uparrow}$ & $\kF a_{\downarrow}$ & $\epsilon_{b}/\epsF$ &
  $ h/\epsF$ & $V_{\uparrow}/\epsF$ & $V_{\downarrow}/\epsF$ & $\Delta\Gamma/\epsF$ \\ \hline
   0.53 & 0.53  & -1.65 &  0.053  &  0.171 & 0.176 & -0.077 \\ \hline
   0.51 & 0.53  & -1.69 &  0.011  &  0.172 & 0.176 & -0.060 \\ \hline
   0.49 & 0.53  & -1.74 & -0.037  &  0.173 & 0.176 & -0.042  \\ \hline
\end{tabular}
 \caption{Values for the AIM model parameters close to the resonance for $a_{\uparrow}$.}
  \label{table:rescond}
\end{table}

The results for these calculation with varying $\kF$ are shown in Fig.~2(a) in
the main text. 
As discussed the bound state peaks lie between $-8\epsF$ and $-2\epsF$. 
Since renormalization effects can shift the energy from
$\epsilon_{b,\sigma}\to\bar\epsilon_{b,\sigma}$, the solution of $\Real G_{b,\sigma}(\omega)^{-1}=0$,
the bound state level can be shifted from their bare values
$\epsilon_{b,\sigma}$ to lower energies. Due to interaction effects the peaks
are broadened, even though they lie outside the continuum.
We see a clear many-body Kondo peak at $\omega=0$. When increasing
$-\epsilon_{b,\sigma}/(\pi \Gamma)$ the width of the resonance does not
decrease very much. The reason for this is that there are two
competing effects. The first is that the bare width
$\pi\Gamma$ is increasing for the set of parameters
($7.8,10.4,17.4$). However, the state for $\kF a_{\sigma}=0.25$ has a
substantially smaller renormalization factor ($z=0.02$ ) than the one for $\kF
a_{\sigma}=0.55$ ($z=0.13$).  Taking into account both of these effects one
understands that the width of the Kondo peak, which is proportional to
$z\Gamma$, does not change very much.

\begin{figure}[htbp]
\centering
\subfigure[]{\includegraphics[width=0.48\columnwidth]{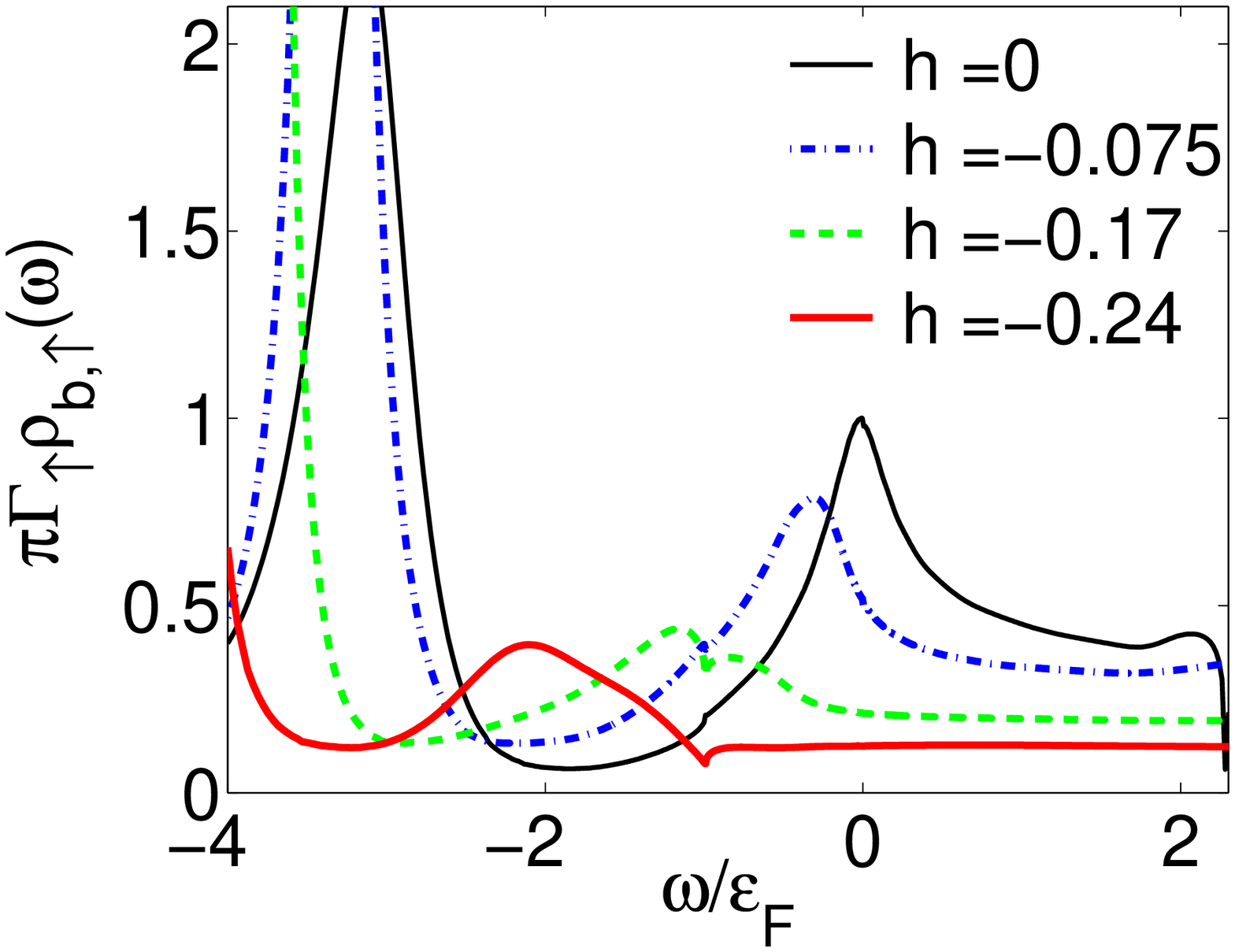}}
\subfigure[]{\includegraphics[width=0.48\columnwidth]{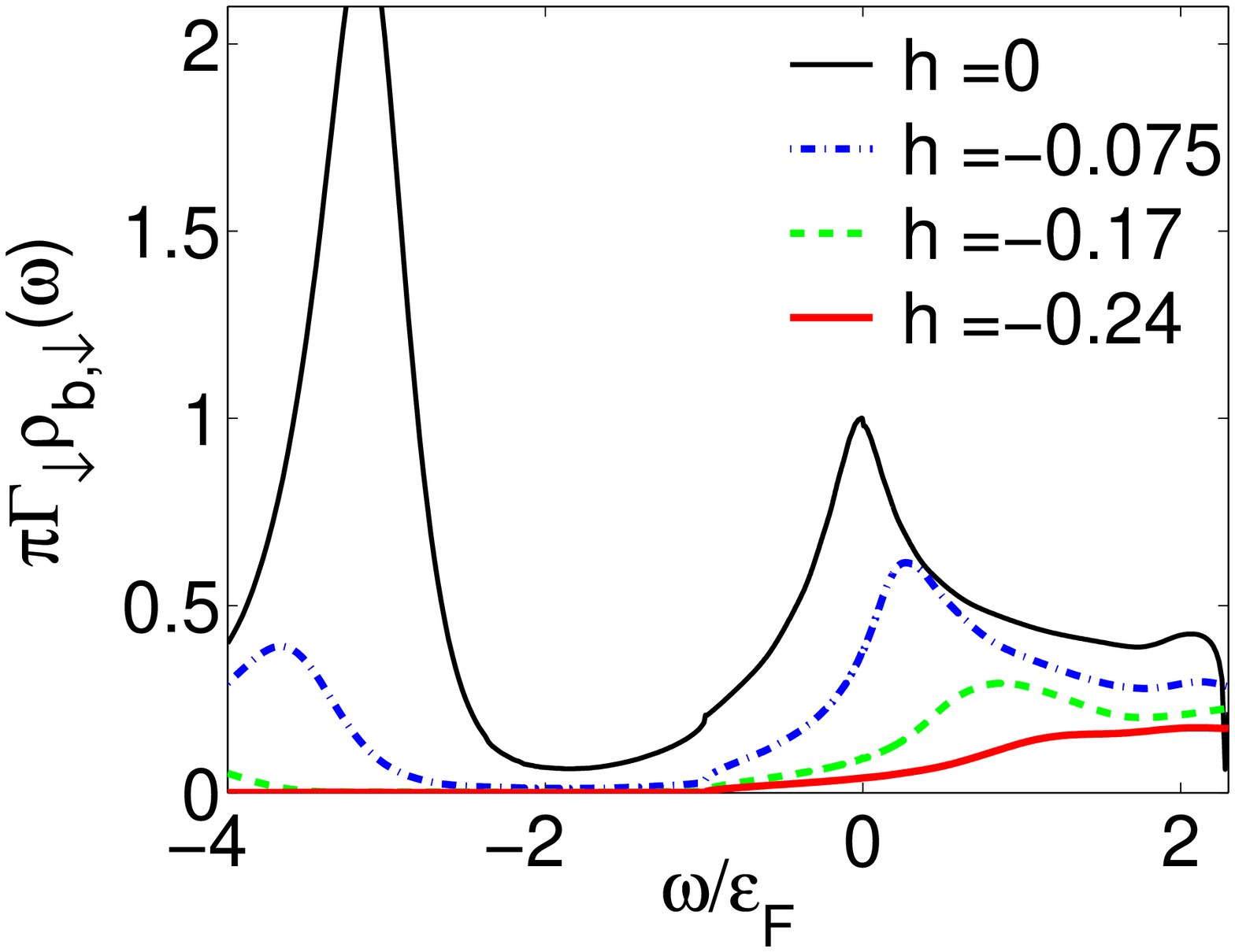}}
\caption{(Color online) Spin resolved spectral functions
  $\rho_{b,\uparrow}(\omega)$ (a) and  $\rho_{b,\downarrow}(\omega)$ (b) for decreasing magnetic field 
  $B=106.25,106.2,106.1,106$G, which yields a number of different values for the
  $\kF a_{\sigma}$ and the field $h$ and the other model parameters (see
  Table~\ref{table:Bdep}), where $\Delta\Gamma\neq 0$. The Kondo resonance 
  is split and suppressed.} 
\label{fig:spec_difB}
\end{figure}

In Table~\ref{table:Bdep} the parameters are shown for the situation where the
field is reduced from $B_s=106.25$G and $\kF=\kF^0$ is held fixed. The
corresponding spectral functions are shown in Fig.~\ref{fig:spec_difB}.

\noindent
The spectral functions show typical behavior of the AIM in a magnetic
field \cite{Hof00,HBK06,BH07a}. We see that $\rho_{b,\uparrow}(\omega)$ gains
spectral weight at $\omega<0$ on decreasing the field $h$, and the Kondo peak
is shifted to negative energies. In contrast, there is a reduction of spectral weight for
$\omega<0$  in $\rho_{b,\downarrow}(\omega)$ when $h$ becomes larger. At the
same time the Kondo peak is shifted to positive energies and broadened. Note
that $\Delta\Gamma<0$ can reduce the magnetic field effect.

\begin{figure}[!htbp]
\centering
\subfigure[]{\includegraphics[width=0.49\columnwidth]{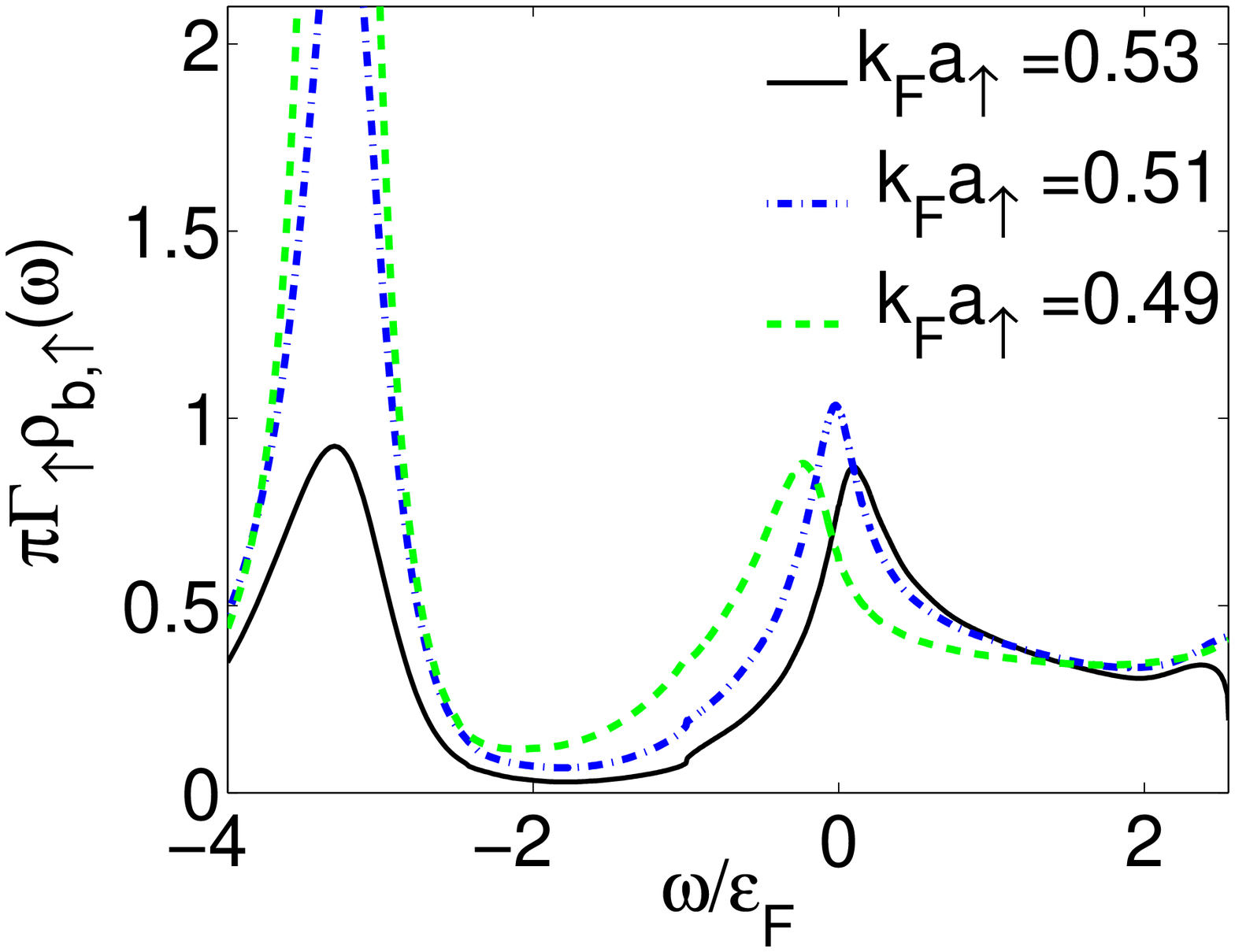}}
\subfigure[]{\includegraphics[width=0.49\columnwidth]{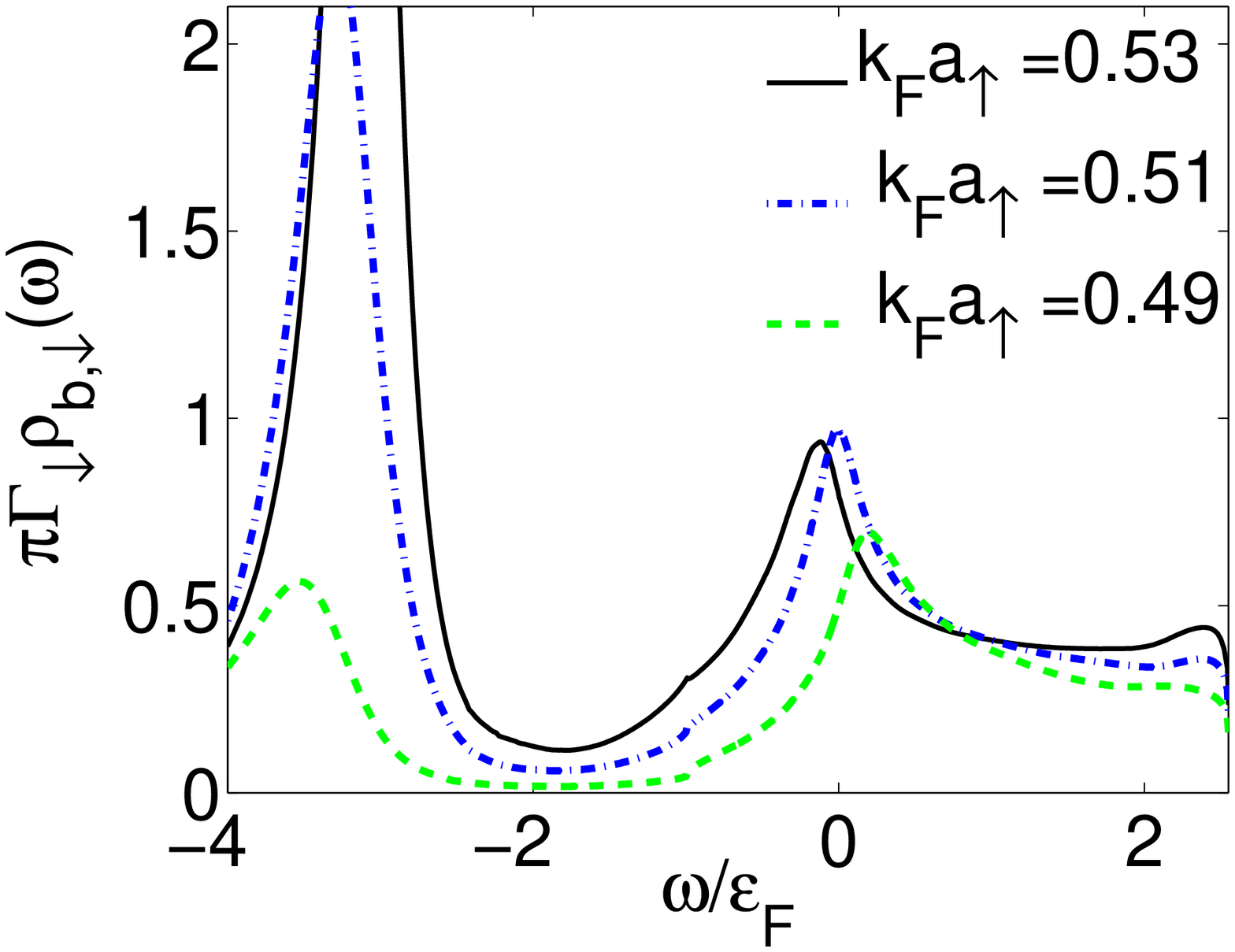}}
\vspace*{-0.3cm}
\caption{(Color online) Spectral functions $\rho_{b,\uparrow}(\omega)$  and
  (b) $\rho_{b,\downarrow}(\omega)$  close to a resonance for a number of
  different values of $\kF a_{\uparrow}$ and the complete set of AIM
  parameters is given in Table~\ref{table:rescond}.}
\label{fig:spec_rescond}
\end{figure}
\noindent

 For illustration  we also discuss a case where the additional resonances are used to tune into
the Kondo-correlated state (cf. Fig.~\ref{fig:a0Bdep}).
We analyze the K-Na system and use $B=106.2$G$<B_s$, where from
Eq.~(3) $\kF a_{\downarrow}=0.475$ and $\kF
a_{\downarrow}=0.53$. Then we tune $\kF 
a_{\uparrow}$ with the additional resonance as discussed in
Fig.~\ref{fig:a0Bdep}(b,c) to satisfy conditions (I-III). 
The model parameters are given in Table~\ref{table:rescond}. We have for $B_{\rm
  K}=106.2$G and $a_{\rm ho}=3.12 \cdot 10^3a_{\rm  B}$, the closest resonance
at $n_{\rm K}\approx 6$. We could also employ the resonance at a different value $n_{\rm K}$
by suitably adjusting $\omega_{\rm ho}$ and hence $a_{\rm ho}$.  
We have assumed that $\kF a_{\downarrow}=0.53$, $\kF r_{e,\uparrow}=0.52$,
$\kF r_{e,\downarrow}=0.47$  are held constant and only $\kF a_{\uparrow}$ varies
close to the resonance. Similar tuning is
possible for other values of $B$ and $a_{\rm ho}$. Generically, for $a_{0,\uparrow}<a_{0,\downarrow}$ one has
$|r_{e,\uparrow}|>|r_{e,\downarrow}|$, which leads to $\Delta\Gamma <0$ and
$h>0$. This leads to the expectation value $n_{\uparrow}-n_{\downarrow}<0$. Now decreasing
$a_{\uparrow}$ close to the resonance from $a_{\uparrow}\simeq a_{\downarrow}$
leads to decreasing $h$, increasing $\Delta\Gamma$, and
$\Lambda_{v,\uparrow}>\Lambda_{v,\downarrow}$. This will lead eventually to a
situation, where the field effect is canceled such that
$n_{\uparrow}-n_{\downarrow}\simeq 0$. Since in this situation still
$V_{\uparrow}\neq V_{\downarrow}$, in general anisotropic Kondo physics is realized. 

The results for the spectral functions are shown in
Fig.~\ref{fig:spec_rescond}.
We identify a clear Kondo peak close to $\omega=0$. The other features are typical
for the Kondo effect in a magnetic field with a slightly shifted resonance and
shifted spectral weight. For $\kF a_{\uparrow}\simeq 0.51 <\kF a_{\downarrow}$
the field effect is roughly canceled and
$\rho_{b,\uparrow}(\omega)\sim\rho_{b,\downarrow}(\omega)$.


\section{S5 RF spectroscopy and experimental signature}
A well-established experimental technique for ultracold atomic gases is radio
frequency (rf) spectroscopy.
The transition rate from a hyperfine state $\ket{\sigma,\vk}$
to a different one $\ket{3,\vk}$, which does not interact with others and is
initially unoccupied, is given by \cite{VPZD11}, 
\begin{equation}
  I_{\sigma}(\omega)=\frac{\Omega^2}{(2\pi)^4}
  \integral{{}^3k}{}{}\!\!\integral{\omega'}{}{}\rho_{\vk,\sigma}(\omega')\rho_{3,\vk}(\omega+\omega') 
  n_{\rm F}(\omega'). 
\label{eq:rfgen}
\end{equation}
$\Omega$ is the intensity and $\omega$ the rf frequency. We assume that
$\ket{3,\vk}$ is a free state shifted in energy by $\omega_{3,\sigma}$ with
respect to the $\ket{\sigma,\vk}$ states, 
$\rho_{3,\vk}(\omega)=\delta(\omega-(\xi_{\vk}+\omega_{3,\sigma}))$. 
The spectral function for $G_{b,\sigma}(\omega)$ in Eq.~(14) reads
 \begin{eqnarray}
&&  \rho_{\vk}(\omega)=   \label{eq:rhok}\\ 
&& \frac{(\eta+n_{\rm imp}V_{\sigma}^2\pi\rho_{b,\sigma}(\omega))/\pi}
{\Big(\omega-\xi_{\vk}- n_{\rm imp}V_{\sigma}^2
  G_{b,\sigma}^R(\omega)\Big)^2\!+\!\Big(\eta+n_{\rm imp}V_{\sigma}^2\pi\rho_{b,\sigma}(\omega)\Big)^2}.    
 \nonumber
\end{eqnarray}
We see that the real part $G_{b,\sigma}^R(\omega)$ can contribute to a shift
of the dispersion and the imaginary part gives an additional broadening.
All terms in Eq.~(\ref{eq:rfgen}) depend on $\vk$ via $\epsilon_{\vk}$ so we
can introduce a density of states (DOS) $\rho_0(\epsilon)$ and write, 
\begin{equation}
  I_{\sigma}(\omega)=\frac{\Omega^2}{2\pi}
  \integral{\epsilon}{}{}\rho_0(\epsilon)\rho_{\epsilon,\sigma}(\epsilon-\mu+\omega_{3,\sigma}-\omega)n_{\rm
    F}(\epsilon-\mu+\omega_{3,\sigma}-\omega). 
\end{equation}
In the limit of low temperature this becomes,
\begin{equation}
  I_{\sigma}(\omega)=\frac{\Omega^2}{2\pi}
  \integral{\epsilon}{-\infty}{\omega-\omega_{3,\sigma}}
\rho_0(\epsilon+\mu)\rho_{\epsilon+\mu,\sigma}(\epsilon+\omega_{3,\sigma}-\omega).  
\end{equation}
In the case of non-interaction fermions,
$\rho_{\epsilon+\mu,\sigma}(\epsilon+\omega_3-\omega)=
\delta(\epsilon+\omega_{3,\sigma}-\omega-\epsilon)=\delta(\omega_{3,\sigma}-\omega)$, so the RF
spectrum reads,  
\begin{equation}
  I_0(\omega)=\frac{2}{3}\frac{\Omega^2}{2\pi}\kF^3 \delta(\omega_{3,\sigma}-\omega).  
\end{equation}

The rf spectra can be understood as the sum of all the transitions from the
modified band dispersion $E_{\vk,\sigma}$ with a certain width
$\Delta_{\vk,\sigma}$ for $\ket{\sigma,\vk}$, to a non-interacting 
dispersion $\epsilon_{\vk}$ shifted by $\omega_{3,\sigma}$ ($\ket{3,\sigma}$). A shift in the dispersion
gives a shift of the rf signal, flattening of the dispersion can give
additional weight at higher or lower energies, and broadening leads to a broadened rf
signal. The Kondo peak on its own leads to a band bending and broadening and
hence a rf peak which is shifted and has high energy tails.

\begin{figure}[!t]
\centering
\includegraphics[width=0.65\columnwidth]{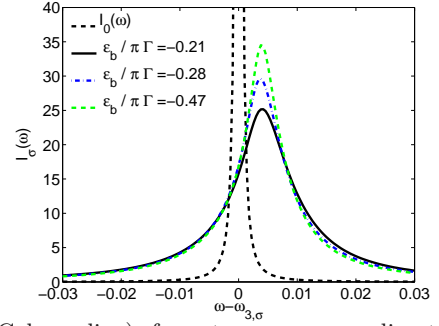}
\vspace*{-0.5cm}
\caption{(Color online) rf spectrum corresponding to
  the spectral function in Fig.~2(a) in the main text and
  the values in Table~\ref{table:kfdep}. We also show $I_0(\omega)$,
  (broadened delta-function) for comparison.}
\label{fig:rfspectrum_difkF}
\vspace*{-0.5cm}
\end{figure}

The rf signal corresponding to Fig.~2(a) in the main text
for an impurity concentration $n_{\rm imp}/N_a\approx 0.03$ is shown in
Fig.~\ref{fig:rfspectrum_difkF}.  

\noindent
When comparing $I_{\sigma}(\omega)$ to the unperturbed signal $I_0(\omega)$ we
find a broadened peak slightly shifted from $\omega=\omega_{3,\sigma}$. This
shape can be traced back to the effect of the coupling of the itinerant states
to the bound state spectral function. In Eq.~(\ref{eq:rhok}) this leads to a
broadening of the spectral 
function and a small band bending close to $\epsilon_{\vk}=\epsF$, where the
Kondo resonance lies. The signal in $I_{\sigma}(\omega)$ narrows for
decreasing Kondo scale due to the reduced modification of $\rho_{\vk,\sigma}(\omega)$. 
The signals in Fig.~\ref{fig:rfspectrum_difkF} are rather characteristic for
the two peak structure in Fig.~2(a), such that the
Kondo-correlated state and position of the Kondo resonance can be identified
well with rf spectroscopy.

\begin{figure}[!htbp]
\centering
\subfigure[]{\includegraphics[width=0.48\columnwidth]{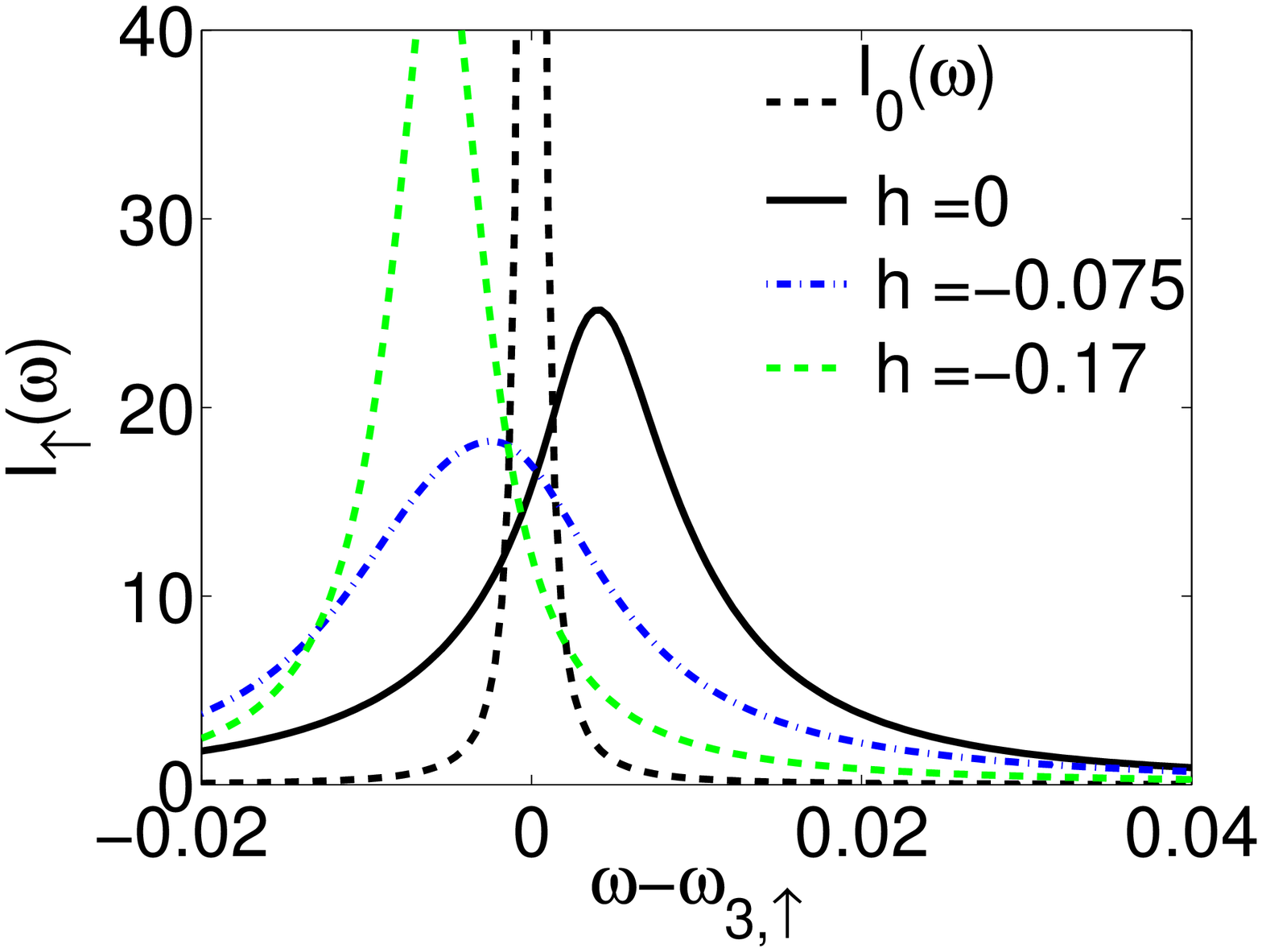}}
\subfigure[]{\includegraphics[width=0.48\columnwidth]{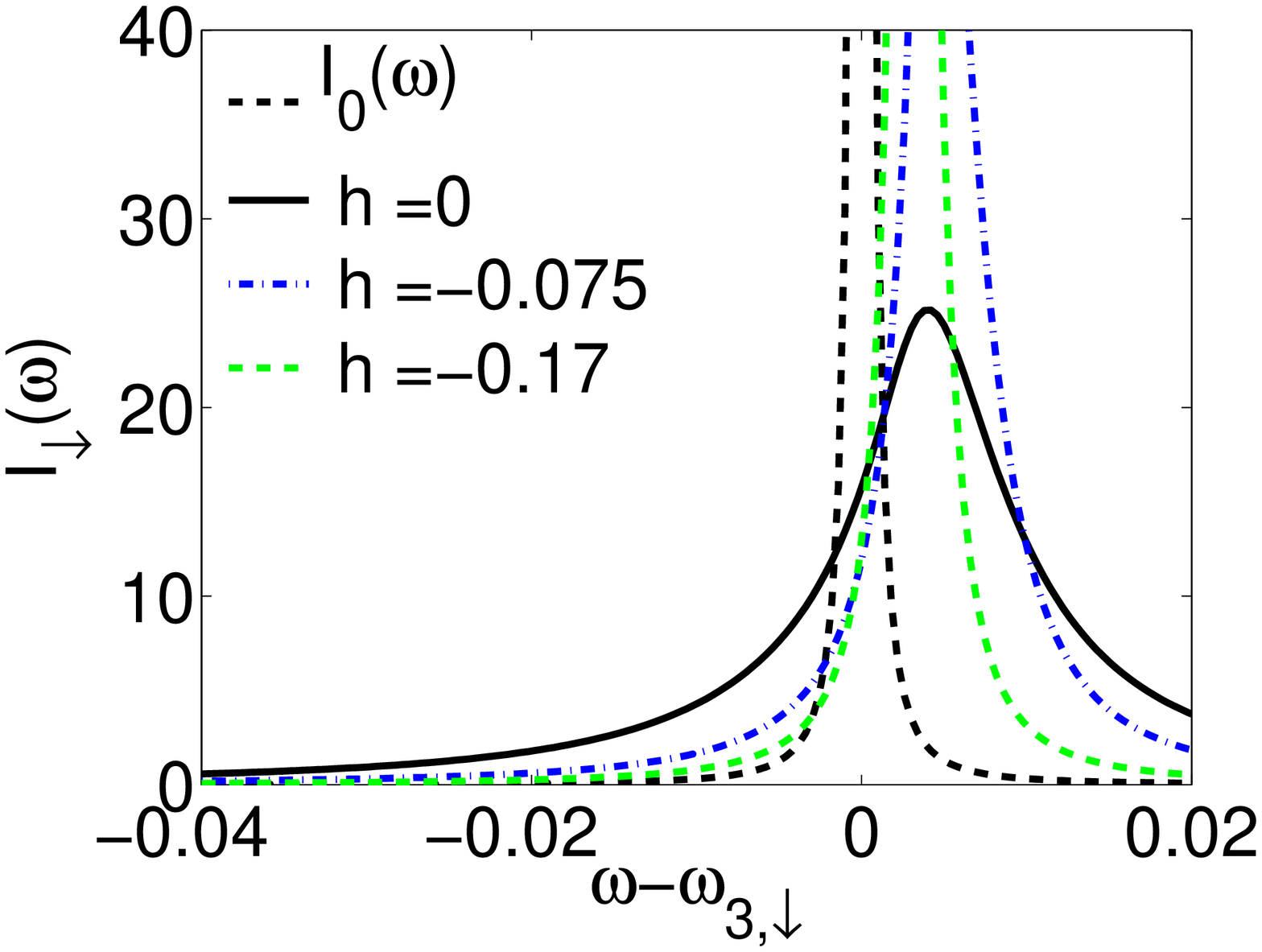}}
\vspace*{-0.5cm}
\caption{(Color online) rf spectra $I_{\uparrow}(\omega)$ (a) and
  $I_{\downarrow}(\omega)$ (b) for decreasing magnetic field $B$ as in
  Fig.~\ref{fig:spec_difB} and model parameters in Table~\ref{table:Bdep}.We also show $I_0(\omega)$,
  (broadened delta-function) for comparison.}
\label{fig:rfspectrum_difB}
\end{figure}

In Fig.~\ref{fig:rfspectrum_difB} we show the rf spectra when varying $B$
corresponding to the results in Fig.~\ref{fig:spec_difB}.

\noindent
The peak in $I_{\uparrow}(\omega)$  is shifted to lower energies. This can be
understood looking at the effect of the moving of the Kondo peak to lower
energies and its effect on $\rho_{\vk,\uparrow}(\omega)$. In contrast, we find that
$I_{\downarrow}(\omega)$ reverts to the non-interacting result 
$I_0(\omega)$ as the $B$ field makes the relevant features disappear in the
spectral function $\rho_{b,\downarrow}(\omega)$ and thus
$\rho_{\vk,\downarrow}(\omega)$. The careful analysis of the $I_{\sigma}(\omega)$
signals as function of magnetic field should be provide a good picture of the tuning
into the Kondo-correlated state.

\begin{figure}[!htbp]
\centering
\subfigure[]{\includegraphics[width=0.49\columnwidth]{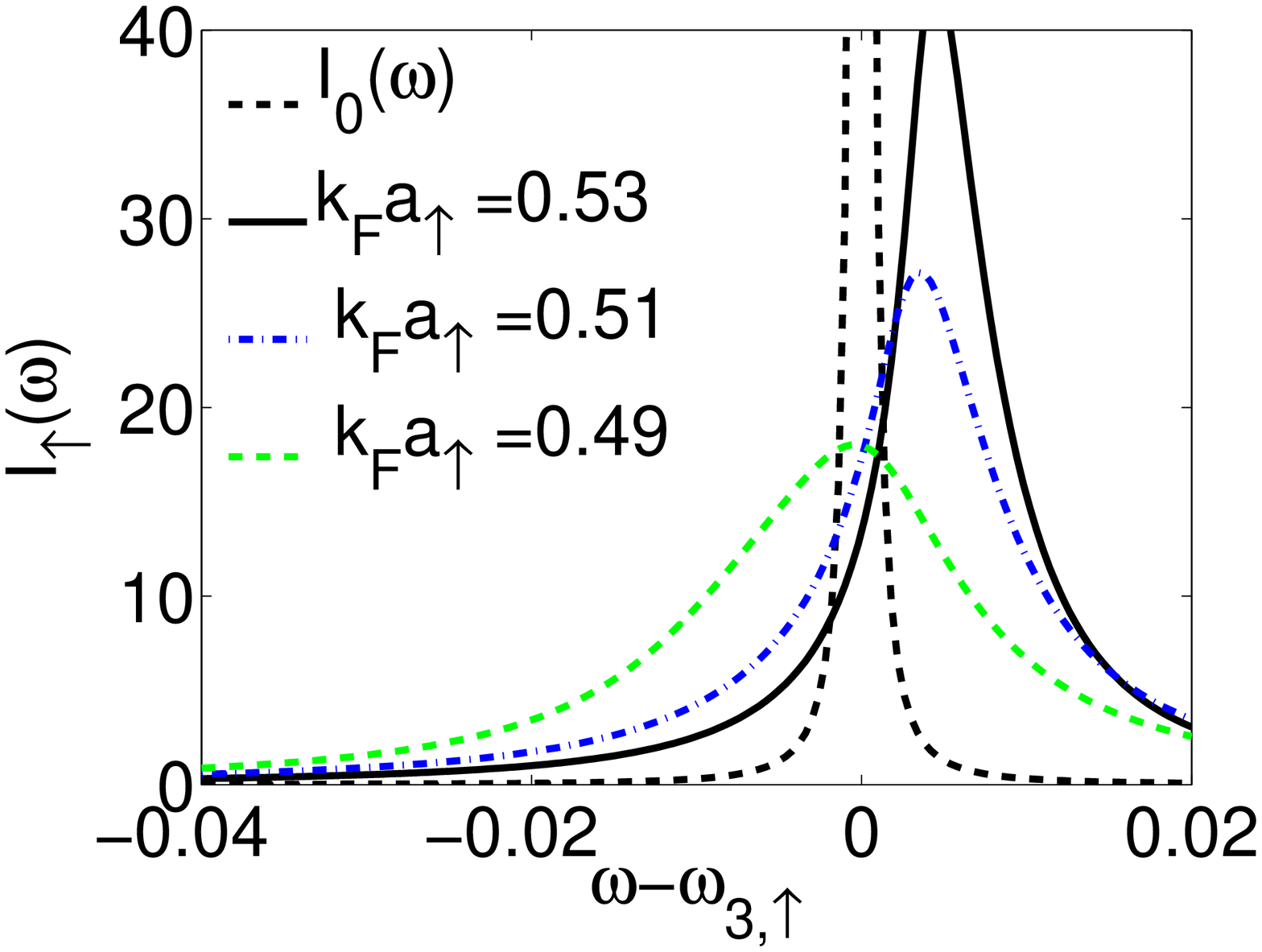}}
\subfigure[]{\includegraphics[width=0.49\columnwidth]{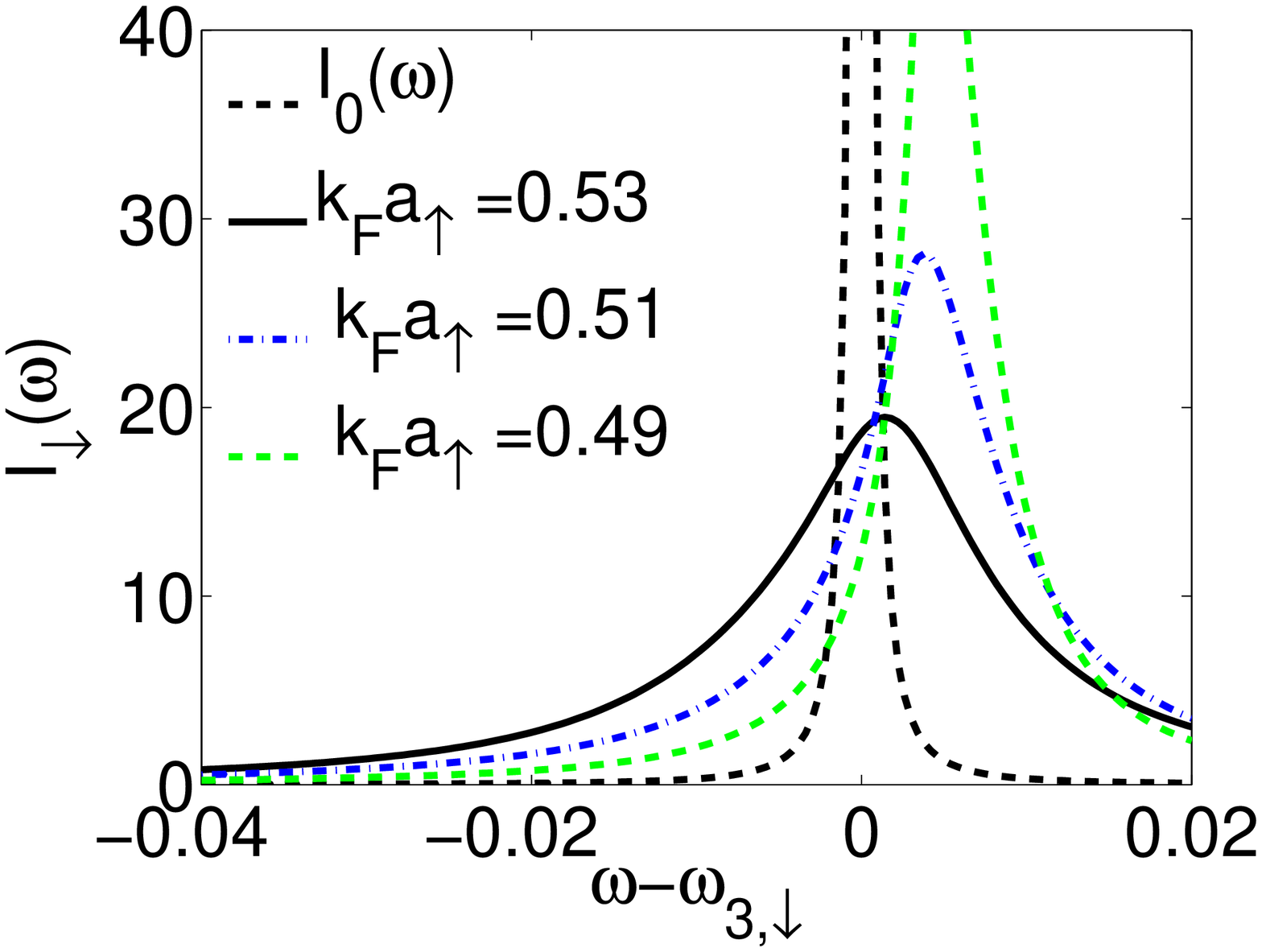}}
\vspace*{-0.5cm}
\caption{(Color online) rf spectra $I_{\uparrow}(\omega)$ (a) and
  $I_{\downarrow}(\omega)$ (b) close 
  to a resonance for a number of different values of $\kF a_{\uparrow}$ as in
  Fig.~\ref{fig:spec_rescond} and the complete set of AIM parameters is given in
  Table~\ref{table:rescond}. We also show $I_0(\omega)$,
  (broadened delta-function) for comparison.}
\label{fig:rfspectrum_rescond} 
\end{figure}

The rf signal corresponding to Fig.~\ref{fig:spec_rescond}(a)(b) is shown in
Fig.~\ref{fig:rfspectrum_rescond}(a)(b). 
When comparing $I_{\sigma}(\omega)$ to $I_0(\omega)$ we find a broadened peak
slightly shifted from $\omega=\omega_{3,\sigma}$. Again we can understand the
shape from the form of the spectral function and its effect on the itinerant
states in Eq.~(\ref{eq:rhok}).  The signal in $I_{\uparrow}(\omega)$ broadens and shifts to
smaller $\omega$ on decreasing $\kF a_{\uparrow}$, and the opposite happens
for $I_{\downarrow}(\omega)$.

\end{appendix}

\bibliography{artikel,biblio1}

\end{document}